\documentclass[11pt,english]{article}
\usepackage[latin9]{inputenc}
\usepackage{amsmath}
\usepackage{amssymb}
\usepackage{graphicx}
\usepackage{setspace}
\usepackage{subscript}
\makeatletter

\providecommand{\tabularnewline}{\\}

\@ifundefined{date}{}{\date{}}
\usepackage{esint}
\usepackage{hyperref}
\usepackage{latexsym}
\usepackage{graphicx}\usepackage{bm}\usepackage{longtable}

\usepackage{xcolor}

\@addtoreset{equation}{section}

\makeatother

\usepackage{babel}
\begin{document}
\title{Three-dimensional Yang-Mills Chern-Simons theory from D3-brane background
with D-instantons}
\maketitle
\begin{center}
Si-wen Li\footnote{Email: siwenli@dlmu.edu.cn}, Sen-kai Luo\footnote{Email: luosenkai@dlmu.edu.cn},
Mu-zhi Tan\footnote{Email: tanmuzhi@dlmu.edu.cn} 
\par\end{center}

\begin{center}
\emph{Department of Physics, School of Science,}\\
\emph{Dalian Maritime University, }\\
\emph{Dalian 116026, China}\\
\par\end{center}

\vspace{8mm}

\begin{abstract}
By constructing the configuration of D3-branes with D(-1)-branes as
D-instantons, we study the three-dimensional Yang-Mills Chern-Simons
theory in holography. Due to the presence of the D-instantons, the
D7-branes with discrepant embedding functions are able to be introduced
in order to include the fundamental fermions (as flavors) and the
Chern-Simons term (at very low energy) in the dual theory. The vacuum
structure at zero temperature is studied in the soliton background
and it illustrates the topological phase transition in the presence
of instantons. Moreover, since the confinement/deconfinement phase
transition could be holographically identified as the Hawking-Page
transition in the bulk, we accordingly calculate the critical temperature
of the deconfinement phase transition by collecting the bulk onshell
action as the thermodynamical free energy. On the other hand, we evaluate
the difference of the entanglement entropy in slab configuration by
using the Ryu-Takayanagi formula since the confinement may also be
characterized by the entanglement entropy. Altogether we find the
behavior of the critical temperature is in qualitative agreement with
the behavior of the critical length determined by the entanglement
entropy which implies the entanglement entropy could indeed be a character
of the confinement in our setup and the D3-D(-1) system would be a
remarkable approach to study the three-dimensional gauge theory.
\end{abstract}
\newpage{}

\tableofcontents{}

\section{Introduction}

In the past decade, a specific class of three-dimensional (3d) Chern-Simons
(CS) theory involving fundamental matters with $N_{f}$ flavors attracts
many interests and the large $N$ 't Hooft limit of such theories
with a $U\left(N\right)$ gauge symmetry has been studied in detail
\cite{key-1,key-2,key-3,key-4,key-5,key-6,key-7}. It is conjectured
there would be a conformal field theory living in the limit of vanishing
flavor mass. Along this direction, some evidence was found which may
be supportable to the conjecture e.g. boson/fermion duality \cite{key-8},
level/rank duality \cite{key-9}. On the other hand, the large $N$
field theory can be analyzed holographically by using string theory
according to gauge/gravity duality or AdS/CFT correspondence which
has become very famous nowadays \cite{key-10,key-11}. Therefore investigation
of CS theory in holography naturally becomes a remarkable topic. And
in this work, our goal is to explore an exactly holographic description
of 3d theory with a CS term.

In string theory, the most simple way to build a holographic duality
for 3d non-supersymmetric theory is to construct the configuration
of coincident $N_{c}$ D3-branes, i.e. to compactify one of the spatial
directions of the D3-brane on a supersymmetry breaking circle $S^{1}$
\cite{key-12,key-13}. Such configuration of $N_{c}$ D3-branes at
large $N_{c}$ limit is described by IIB supergravity and it has been
attempted to study the 3d Yang-Mills Chern-Simons (YMCS) theory or
3d QCD (QCD\textsubscript{3}) in holography e.g. the vacuum structure
\cite{key-14,key-15} and the quantum Hall Effect \cite{key-16}.
However, the D3-brane background does not include dynamical D-instanton
which would play the important role to involve the topological properties
in the dual theory. For example, as it is known the YM instanton in
4d quantum chromodynamics (QCD) is topologically nontrivial excitation
of the vacuum which contributes to the thermodynamics of the theory
and relates to chiral symmetry breaking \cite{key-17,key-18}. Based
on string theory, the D$p$-D($p-4$) brane system ($N$ D$p$-branes
and $M$ D($p-4$)-branes) has gauge theory instantons with exactly
$4NM$ moduli as in gauge theory \cite{a1,a2}, thus, via gauge/gravity
duality, a holographic way to include the YM instanton is to take
into account the configuration of coincident $N_{c}$ D3-branes (choosing
$p=3$) with $N_{\mathrm{D}}$ D(-1)-branes as D-instantons smeared
homogeneously in the worldvolume of the D3-branes \cite{key-19,key-20,key-21}.
The system of D3-branes with dynamical D-instantons succeeds to describe
the 4d QCD-like theories and the presence of D-instantons could reflect
some topological properties of the dual theory \cite{key-22,key-23,key-24,key-25}.
So the D3-D(-1) approach has provided an impressive interpretation
of the D-instanton.

In order to work out a holographic description of 3d theory with a
CS term, in this note we focus on constructing the D3-D-instanton
configuration by following the standard technique \cite{key-12,key-13}
in string theory because the D3-D(-1) approach would turn out that
the D(-1)-branes, as the D-instantons, could reduce to a CS term in
the 3d dual theory. Specifically, since the D(-1)-branes are dynamical
in the supergravity background, there would be a non-vanished Ramond-Ramond
zero form $C_{0}$ coupling to these D(-1)-branes. So when we examine
the dual theory by considering a probe D3-brane, its action would
contain a term as $\int C_{0}F\wedge F$. As we will focus on the
3d dual  theory obtained by compactifying one of the spatial directions
on the circle $S^{1}$, thus below the energy scale of the circle,
the term $\int C_{0}F\wedge F$ can be integrated out to become a
3d CS term as $\int C_{0}F\wedge F\sim\int dC_{0}\wedge\omega_{3}\sim\frac{k_{b}}{4\pi}\mathrm{Tr}\int\omega_{3}$
where $\omega_{3}$ refers to the CS 3-form and $k_{b}$ corresponds
to the boundary value of $C_{0}$. Afterwards once the soliton AdS
solution is chosen, it will lead to an area law due to the asymptotics
of the Wilson loop, so confinement is also  expected to exhibit in
the dual theory at low temperature. To further include matters, the
flavors are introduced by embedding a stack of probe D7-branes \cite{key-26}
and the spontaneous breaking of chiral symmetry can be identified
as the separation of $N_{f}$ flavor branes in IR region depicted
by the IIB supergravity as the holographic description of \cite{key-27}.
Moreover, additional D7-branes as CS branes with a discrepancy embedding
function can be put into the background due to the presence of the
D-instantons, accordingly at very low energy, the dual theory could
be a pure CS theory. We specifically analyze the behaviors of the
embedding functions of the various D7-branes and evaluate the associated
free energy density by including both flavors and CS term. The result
illustrates the topological phase transition which is enhanced by
the D-instantons and it seems to be qualitatively consistent with
that the presence of CS term topologically contributes to the mass
parameter \cite{key-28} and the D0-D4 approach in 4d \cite{key-29,key-30,key-31,key-32}.

Besides, we evaluate the critical temperature of the deconfinement
phase transition in this holographic setup since the dual theory is
expected to exhibit the confinement. While the deconfining geometry
in holography is less clear, the phase transition at a critical temperature
can be identified as the bubble/black brane transition, or namely
Hawking-Page transition, in the bulk which is very suggestive of the
deconfinement transition \cite{key-33,key-34,key-35,key-36}. Following
this idea, we collect the total onshell action as the holographic
free energy consisted of the bulk fields (as the color sector), the
flavor and the CS branes, in the soliton and black brane background
respectively. By comparing the free energy, we find the charge density
of D-instantons indeed contributes to the thermodynamics and the critical
temperature is decreased by the D-instantons through the flavor and
CS branes. However, at large-$N_{c}$ limit, the critical temperature
becomes independent on the D-instantons since the backreaction of
the CS branes is not included in this setup. To parallel examine whether
our analysis for the deconfinement transition is consistent, we on
the other side calculate the entanglement entropy holographically
in this system since the entanglement entropy may also be a characteristic
tool to detect the confinement in the dual theory \cite{key-37,key-38,key-39,key-40}.
Using the Ryu-Takayanagi (RT) formula \cite{key-41} with the slab
geometry, the holographic entanglement entropy can be obtained and
it exhibits a first order phase transition at a critical length who
behaves similarly as the critical temperature evaluated by the thermodynamics.
In this sense, we believe the configuration of D3-branes with D-instantons
would be a remarkable approach to study 3d gauge theory.

The outline of this note is as follows. In Section 2, we construct
the black D3-D(-1) solution to obtain a confining geometry and examine
the dual theory by a probe D3-brane. Afterwards, we analyze the embedding
function of the flavor and CS branes, compute the free energy by including
the flavor and CS term in Section 3 then evaluate the vacuum structure
of the dual theory in Section 4. In Section 5, we investigate the
deconfinement phase transition by comparing the free energy of this
model thermodynamically and compute the variation of the entanglement
entropy as a parallel verification. Summary and comments are given
in the last section.

\section{Three-dimensional theory from confining geometry}

In this section, we will briefly review the system of $N_{c}$ D3-branes
with $N_{\mathrm{D}}$ D-instantons i.e. the D(-1)-branes, then construct
the background geometry for a confining dual theory at large-$N_{c}$
limit. 

The D3-D(-1) brane system is geometrically represented by a deformed
D3-brane solution with a nontrivial Ramond-Ramond (R-R) scalar field
$C_{0}$ which is recognized as a marginal \textquotedblleft bound
state\textquotedblright{} of D3-branes with $N_{\mathrm{D}}$ smeared
D(-1)-branes. We denote the $N_{c}$ D3-branes as color branes. In
the large $N_{c}$ limit, the 10 dimensional (10d) type IIB supergravity
action, as the effective action, describes the low-energy dynamics
of this system which in string frame is given as, 

\begin{equation}
S_{\mathrm{IIB}}=\frac{1}{2\kappa_{10}^{2}}\int d^{10}x\sqrt{-g}\left[e^{-2\Phi}\left(\mathcal{R}+4\partial\Phi\cdot\partial\Phi\right)-\frac{1}{2}\left|F_{1}\right|^{2}-\frac{1}{2}\left|F_{5}\right|^{2}\right].\label{eq:1}
\end{equation}
Here $2\kappa_{10}^{2}=\left(2\pi\right)^{7}l_{s}^{8}$ is the 10d
gravity coupling constant, $l_{s},g_{s}$ is respectively the length
and the coupling constant of string and $F_{1,5}$ is the field strength
of the R-R zero and four form $C_{0,4}$. The near-horizon solution
of non-extremal D3-branes with a non-trivial $C_{0}$ in string frame
reads,

\begin{align}
ds^{2} & =e^{\frac{\phi}{2}}\left\{ \frac{r^{2}}{R^{2}}\left[-f_{T}\left(r\right)dt^{2}+d\mathbf{x}\cdot d\mathbf{x}\right]+\frac{1}{f\left(r\right)}\frac{R^{2}}{r^{2}}dr^{2}+R^{2}d\Omega_{5}^{2}\right\} ,\nonumber \\
e^{\phi} & =1+\frac{Q}{r_{H}^{4}}\ln\frac{1}{f\left(r\right)},\ f_{T}\left(r\right)=1-\frac{r_{H}^{4}}{r^{4}},\ F_{5}=dC_{4}=g_{s}^{-1}\mathcal{Q}_{3}\epsilon_{5},\nonumber \\
F_{1} & =dC_{0},\ C_{0}=-ie^{-\phi}+i\chi,\ \phi=\Phi-\Phi_{0},\ e^{\Phi_{0}}=g_{s},\label{eq:2}
\end{align}
where $\epsilon_{5}$ is the volume element of a unit $S^{5}$ and

\begin{equation}
R^{4}=4\pi g_{s}N_{c}l_{s}^{4},\ \mathcal{Q}_{3}=4R^{4},\ Q=\frac{N_{\mathrm{D}}}{N_{c}}\frac{\left(2\pi\right)^{4}\alpha^{\prime2}}{V_{4}}\mathcal{Q}_{3}.
\end{equation}
This solution describes that the D-instanton charge $N_{\mathrm{D}}$
is smeared over the worldvolume $V_{4}$ of the coincident black $N_{c}$
D3-branes homogeneously with a horizon at $r=r_{H}$. And it implies
$N_{\mathrm{D}}/N_{c}$ must be fixed since the backreaction of the
D-instantons has been involved in the bulk action. The dual theory
of this system is conjectured as the 4d $\mathcal{N}=4$ super Yang-Mills
theory (SYM) in a self-dual gauge field background or with a dynamical
axion at finite temperature characterized by the order parameter $Q$.
In order to obtain a confining or QCD-like dual theory, let us follow
the discussion in \cite{key-12,key-13}. Specifically we first take
one of the three spatial dimensions $x^{i}$ of the D3-branes to be
compactified on a circle $S^{1}$ with a period $x^{i}\sim x^{i}+\delta x^{i}$.
Hence below the Kaluza-Klein energy scale defined as $M_{KK}=2\pi/\delta x^{i}$,
the dual theory becomes effectively three-dimensional. Then we are
going to get rid of all massless particles other than the gauge fields.
The most simple way to achieve this is to impose the anti-periodic
and periodic boundary condition on fermion and bosonic fields respectively
along $S^{1}$. So the supersymmetric fermions and scalars in the
dual theory acquire mass of order $M_{KK}$ which is accordingly decoupled
in the low-energy dynamics. Next we perform a double Wick rotation
on the D(-1)-D3 brane background i.e. $t\rightarrow-ix^{i},x^{i}\rightarrow-it$
to identify the bulk gravity solution as its holographic correspondence.
Without loss of generality, let us denote the direction along $S^{1}$
as $x^{i}=x^{3}$ throughout this manuscript, thus the confining solution
of non-extremal D3-branes with smeared D-instantons is obtained as,

\begin{align}
ds^{2} & =e^{\phi/2}\left\{ \frac{r^{2}}{R^{2}}\left[\eta_{ab}dx^{a}dx^{b}+f\left(r\right)\left(dx^{3}\right)^{2}\right]+\frac{1}{f\left(r\right)}\frac{R^{2}}{r^{2}}dr^{2}+R^{2}d\Omega_{5}^{2}\right\} ,\nonumber \\
f\left(r\right) & =1-\frac{r_{KK}^{4}}{r^{4}},\ a,b=0,1,2,\label{eq:5}
\end{align}
where the solution of dilaton $\Phi$ and R-R fields $C_{0,4}$ remains.
Since the warp factor $e^{\phi/2}\frac{r^{2}}{R^{2}}$ never goes
to zero, the solution (\ref{eq:5}) defined for $r>r_{KK}$ does not
have a horizon. And it would lead to an area law in the dual theory
according to the asymptotics of the Wilson loop in this geometry.
Namely below the energy scale $M_{KK}$, the dual field theory should
exhibit confinement. To avoid the conical singularities in the region
of $r>r_{KK}$, we have to further require,

\begin{equation}
M_{KK}=\frac{2r_{KK}}{R^{2}}.
\end{equation}
Afterwards, the dual theory can be examined by taking into account
the action of a probe D3-brane which is expected to be a 3d Yang-Mills
(YM) plus Chern-Simons (CS) theory at $r\rightarrow\infty$ as,

\begin{align}
S_{\mathrm{D3}} & =-\mu_{3}\int d^{4}xe^{-\phi}\mathrm{Str}\sqrt{-\det\left(g+\mathcal{F}\right)}+\mu_{3}\int C_{4}+\frac{1}{2}\mu_{3}\mathrm{Tr}\int C_{0}\mathcal{F}\land\mathcal{F}\nonumber \\
 & \simeq-\frac{1}{2g_{YM}^{2}}\mathrm{Tr}\int d^{4}xF_{\mu\nu}F^{\mu\nu}-\frac{1}{4\pi}\mathrm{Tr}\int dC_{0}\wedge\omega_{3}+\mathcal{O}\left(F^{4}\right)\nonumber \\
 & =-\frac{1}{2g_{3d,YM}^{2}}\mathrm{Tr}\int d^{3}xF_{ab}F^{ab}+i\frac{k_{b}}{4\pi}\mathrm{Tr}\int_{\mathbb{R}^{1,2}}\omega_{3},\label{eq:6}
\end{align}
where $\mathcal{F}=2\pi\alpha^{\prime}F$ is the gauge strength, $\mu_{p}=\left(2\pi\right)^{-p}l_{s}^{-p-1}$
refers to the D-brane charge and $\omega_{3}$ is the Chern-Simons
3-form,

\begin{align}
\omega_{3} & =A\wedge dA+\frac{2}{3}A\wedge A\wedge A.
\end{align}
By imposing the background solution, it leads to

\begin{equation}
dC_{0}\big|{}_{r\rightarrow\infty}=-ik_{b}M_{KK}\delta\left(x^{3}-\bar{x}^{3}\right)dx^{3},g_{3d,YM}^{2}=\frac{g_{YM}^{2}M_{KK}}{2\pi},
\end{equation}
where we have assumed that $\omega_{3}$ is independent on $x^{3}$
and does not have components along $x^{3}$. So (\ref{eq:6}) represents
the YM-CS action located at $x^{3}=\bar{x}^{3}$, which means $C_{0}\big|_{r\rightarrow\infty}=0$
if $x^{3}\in\left(0,\bar{x}^{3}\right)$; $C_{0}\big|_{r\rightarrow\infty}\neq0$
if $x^{3}\in\left(\bar{x}^{3},2\pi M_{KK}^{-1}\right)$. In this case,
we have to slightly modify the supergravity solution for $C_{0}$
in (\ref{eq:2}) as,

\begin{equation}
\chi=1-k_{b}M_{KK}\Theta\left(x^{3}-\bar{x}^{3}\right),
\end{equation}
where $\Theta\left(x^{3}-\bar{x}^{3}\right)$ is the step function.

\section{Flavor and Chern-Simons brane}

In this section, let us discuss the embedding of flavor and CS brane
in the D3-brane background with D-instantons (\ref{eq:5}) in holography. 

\subsection{Adding flavors}

According to the dictionary of AdS/CFT, introducing flavors into the
holographic background is to add fundamental matter in the dual theory
\cite{key-26}. So follow the discussion of D3/D7 approach, we put
a stack of $N_{f}$ D7-branes as probes, as $N_{f}$ copies of fundamental
flavors, into our background (\ref{eq:5}) and the configuration of
various D-branes is illustrated in Table \ref{tab:1}. 
\begin{table}
\begin{centering}
\begin{tabular}{|c|c|c|c|c|c|c|c|c|c|c|c|}
\hline 
 & -1 & 0 & 1 & 2 & (3) & 4 (r) & 5 & 6 & 7 & 8 & 9\tabularnewline
\hline 
\hline 
D(-1)-branes & - &  &  &  &  &  &  &  &  &  & \tabularnewline
\hline 
Color D3-branes &  & - & - & - & - &  &  &  &  &  & \tabularnewline
\hline 
Flavor D7-branes &  & - & - & - &  & - & - & - & - & - & \tabularnewline
\hline 
CS D7-branes &  & - & - & - &  &  & - & - & - & - & -\tabularnewline
\hline 
\end{tabular}
\par\end{centering}
\caption{\label{tab:1} The configuration of various D-branes. ``-'' represents
the D-branes extend along this direction. Note that ``-1'' is vertical
to all the directions of bulk spacetime.}
\end{table}
 Note that in this configuration the supersymmetry is completely broken
even in an extremal D3-brane background since the leftover direction
$x^{9}$ is transverse to both flavor D7- and color D3-branes which
leads to 6 mixed Neumann-Dirichlet boundary conditions. The bare mass
of flavors corresponds to the separation between color and flavor
branes at the UV boundary, which means the scalar field respected
to $x^{9}$ on the worldvolume of the D7-branes is the mass operator
$\bar{\psi}\psi$ in the dual field theory. 

Since the directions $x^{4}...x^{9}$ transverse to the $N_{c}$ D3-branes
are usually described by spherical coordinates , for convenience we
introduce a new radius coordinate $\rho$ as,

\begin{equation}
r\left(\rho\right)=\left(\rho^{2}+\frac{r_{KK}^{4}}{4\rho^{2}}\right)^{1/2},\rho>\frac{r_{KK}}{\sqrt{2}},
\end{equation}
thus the metric (\ref{eq:5}) on coordinate $\rho$ can be written
as,

\begin{equation}
ds^{2}=e^{\phi/2}\left\{ \frac{r^{2}}{R^{2}}\left[\eta_{\alpha\beta}dx^{\alpha}dx^{\beta}+f\left(r\right)\left(dx^{3}\right)^{2}\right]+\frac{R^{2}}{\rho^{2}}\left(d\rho^{2}+\rho^{2}d\Omega_{5}^{2}\right)\right\} .
\end{equation}
Then let us choose the spherical coordinates $\lambda,\Omega_{4}$
to reparametrize the directions $x^{4}...x^{8}$ which are part of
the worldvolume of flavor branes and separate transverse coordinate
$x^{9}\equiv u$ with $\rho^{2}=\lambda^{2}+u^{2}$. Afterwards the
metric on $\left\{ x^{a},x^{3},\lambda,\Omega_{4},u\right\} $ takes
the form as,

\begin{equation}
ds^{2}=e^{\phi/2}\left\{ \frac{r^{2}}{R^{2}}\left[\eta_{ab}dx^{a}dx^{b}+f\left(r\right)\left(dx^{3}\right)^{2}\right]+\frac{R^{2}}{\rho^{2}}\left(d\lambda^{2}+\lambda^{2}d\Omega_{4}^{2}+du^{2}\right)\right\} ,
\end{equation}
where $r=r\left(\rho\right)$. Embedding the flavor brane into $\left\{ x^{a},\lambda,\Omega_{4}\right\} $
at a constant $x^{3}$ with $u=u\left(\lambda\right)$, the induced
metric on the flavor D7-brane becomes,

\begin{equation}
ds_{\mathrm{D}7}^{2}=e^{\phi/2}\left\{ \frac{r^{2}}{R^{2}}\eta_{\alpha\beta}dx^{\alpha}dx^{\beta}+\frac{R^{2}}{\rho^{2}}\left[\left(1+u^{\prime2}\right)d\lambda^{2}+\lambda^{2}d\Omega_{4}^{2}\right]\right\} .
\end{equation}
Note that we need to impose the following boundary condition ($\frac{du}{d\lambda}\equiv u^{\prime}$),

\begin{equation}
u^{\prime}\big|_{\lambda=0}=0,u\big|_{\lambda=\lambda_{\infty}}=u_{\infty}.\label{eq:14}
\end{equation}
We have use $\lambda_{\infty}$ to denote the UV boundary or UV cutoff
in the dual field theory. So for a single D7-brane, its action can
be collected as,

\begin{equation}
S_{\mathrm{D}7}=-T_{\mathrm{D}7}\int d^{8}xe^{-\phi}\sqrt{-g_{\mathrm{D7}}},\label{eq:15}
\end{equation}
where $T_{\mathrm{D}7}=g_{s}^{-1}\mu_{p}$ is the tension of the Dp-brane.
Plugging the solution (\ref{eq:2}) into (\ref{eq:15}), the action
of a probe flavor brane is obtained as,

\begin{equation}
S_{\mathrm{D}7}=-T_{\mathrm{D}7}V_{3}V_{S^{4}}R^{2}\int d\lambda e^{\phi\left(\rho\right)}\left(\rho^{2}+\frac{r_{KK}^{4}}{4\rho^{2}}\right)^{3/2}\frac{\lambda^{4}}{\rho^{5}}\sqrt{1+u^{\prime2}},
\end{equation}
where $V_{3},V_{S^{4}}$ refers to the Minkowskian worldvolume of
D3-brane and the volume of a unit $S_{4}$. By varying the D7-brane
action respected to $u\left(\lambda\right)$, the associated equation
of motion is, 

\begin{align}
 & \frac{d}{d\lambda}\left[e^{\phi}\left(r_{KK}^{4}+4\rho^{4}\right)^{3/2}\frac{\lambda^{4}}{8\rho^{8}}\frac{u^{\prime}}{\sqrt{1+u^{\prime2}}}\right]\nonumber \\
= & -e^{\phi}\left[r_{KK}^{4}+\rho^{4}-\frac{1}{8}\rho\frac{d\Phi}{d\rho}\left(r_{KK}^{4}+4\rho^{4}\right)\right]\left(r_{KK}^{4}+4\rho^{4}\right)^{1/2}\frac{\lambda^{4}u}{\rho^{10}}\sqrt{1+u^{\prime2}}.\label{eq:17}
\end{align}
In order to obtain the embedding function $u\left(\lambda\right)$,
we have to solve (\ref{eq:17}) with (\ref{eq:14}). So let us analyze
massless and massive embedding of the flavor brane respectively. 

\subsubsection*{Massless case}

First let us investigate the case of the limit $r_{KK}\rightarrow0$\footnote{Since the value (of e.g. dilaton) at $r_{KK}=0$ may be divided, we
will not take the strict limit although the limit of $r_{KK}\rightarrow0$
is well defined in the D-brane background \cite{key-12}. An effective
way is to choose $r_{KK}=\varepsilon$ where $\varepsilon$ is infinitesimally
small, then take $\varepsilon\rightarrow0$ in the final result, so
there would be no inconsistency in our calculation.} which corresponds to the extremal D3-D(-1) solution. The equation
of motion (\ref{eq:17}) comes to

\begin{equation}
\frac{d}{d\lambda}\left[e^{\phi}\frac{\lambda^{4}}{\rho^{2}}\frac{u^{\prime}}{\sqrt{1+u^{\prime2}}}\right]=-\frac{2\lambda^{4}u}{\rho^{4}}e^{\phi}\left(1-\frac{1}{2}\rho\frac{d\phi}{d\rho}\right)\sqrt{1+u^{\prime2}},\label{eq:18}
\end{equation}
where

\begin{equation}
\phi\rightarrow1+\frac{Q}{r^{4}},
\end{equation}
in the limit of $r_{KK}\rightarrow0$. It is clear that at $\lambda=0$
the right-hand side of (\ref{eq:18}) is not vanished unless $u\left(\lambda\right)=0$
is the solution. We expect $u\left(\lambda\right)=0$ to be an unstable
solution as it is discussed in the D3/D7 approach \cite{key-14,key-15}
since this solution is invariant under the parity transformation $u\left(\lambda\right)\rightarrow-u\left(\lambda\right)$.

\begin{figure}
\begin{centering}
\includegraphics[scale=0.5]{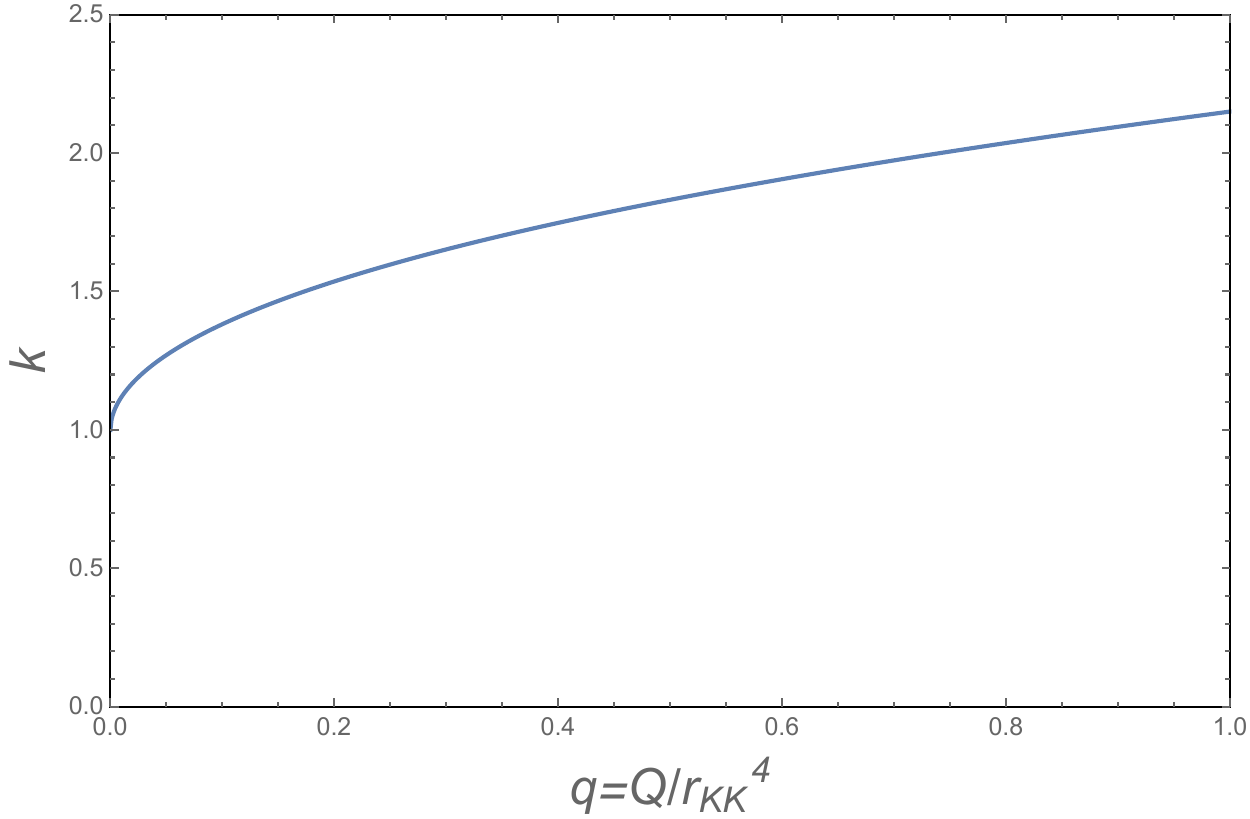}
\par\end{centering}
\caption{\label{fig:1} Numerical solution of $k$ and $q$.}
\end{figure}

\begin{figure}
\begin{centering}
\includegraphics[scale=0.5]{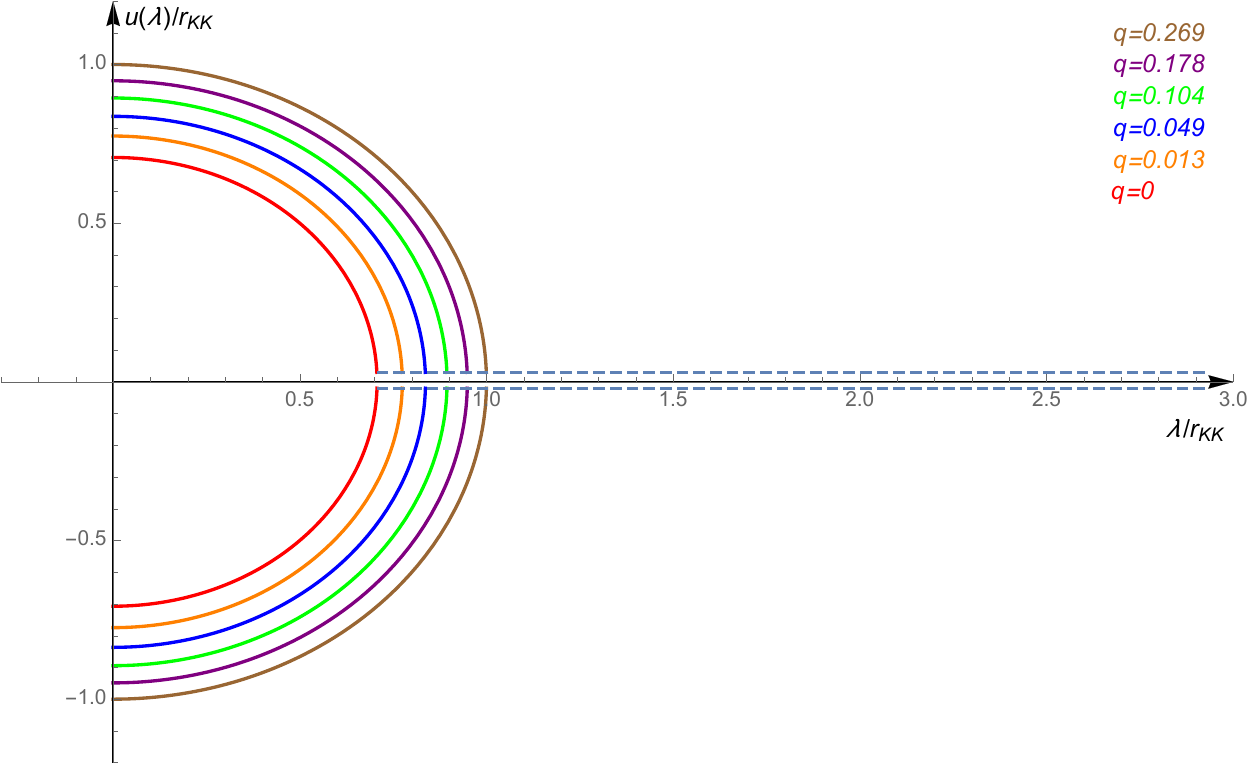}
\par\end{centering}
\caption{\label{fig:2} The maximal embedding of flavor D7-branes in the $u,\lambda$
plane with various $q$.}
\end{figure}

Then let us investigate the case of $r_{KK}>0$. In the massless,
we need to choose $u_{\infty}=0$ in (\ref{eq:14}) since there is
a parity transformation $u\left(\lambda\right)\rightarrow-u\left(\lambda\right)$
in the massless case and $u_{\infty}$ corresponds to the bare mass
of the flavors. In order to search for an analytical solution, we
use the following ansatz for $u\left(\lambda\right)$ as,

\begin{equation}
u\left(\lambda\right)=\begin{cases}
\pm\sqrt{\frac{r_{KK}^{2}}{2}k-\lambda^{2}}, & 0\leq\lambda\leq\sqrt{\frac{k}{2}}r_{KK},\\
0, & \lambda>\sqrt{\frac{k}{2}}r_{KK},
\end{cases}\label{eq:20}
\end{equation}
where $k\equiv k\left(q\right)$ is a constant dependent on $q=Q/r_{KK}^{4}$
only. Notice that $r\in\left(r_{KK},\infty\right)$ so that $k\geq1$.
Plugging (\ref{eq:20}) into (\ref{eq:17}), it leads to a constraint
equation which determines the relation of $k$ and $q$ as,

\begin{equation}
3+3k^{4}-2k^{2}\left(3+8q\right)-3q\left(k^{2}-1\right)^{2}\ln\left[\left(\frac{k^{2}-1}{k^{2}+1}\right)^{2}\right]=0.
\end{equation}
This equation can be numerically solved and the relation of $k$ and
$q$ is illustrated as in Figure \ref{fig:1}. 
 The solution (\ref{eq:20}) has two branches which refers to a pair
of D7-branes wrapping the upper and lower half-five-sphere with various
numbers of D-instantons represented by $q$. The flavor branes wrapping
the upper and lower half-five-sphere have opposite parity as it is
illustrated in Figure \ref{fig:2}. 
 Since the relation of $k$ and $q$ smoothly returns to the case
without D-instantons i.e. $k=1,q=0$, the solution (\ref{eq:20})
corresponds to the maximal embedding of the flavor branes as it is
in the D3/D7 approach which refers to the embedded flavor branes have
maximal energy among all possible solutions to (\ref{eq:17}) with
$u_{\infty}=0$ in the boundary condition (\ref{eq:14}). 

\begin{figure}
\begin{centering}
\includegraphics[scale=0.5]{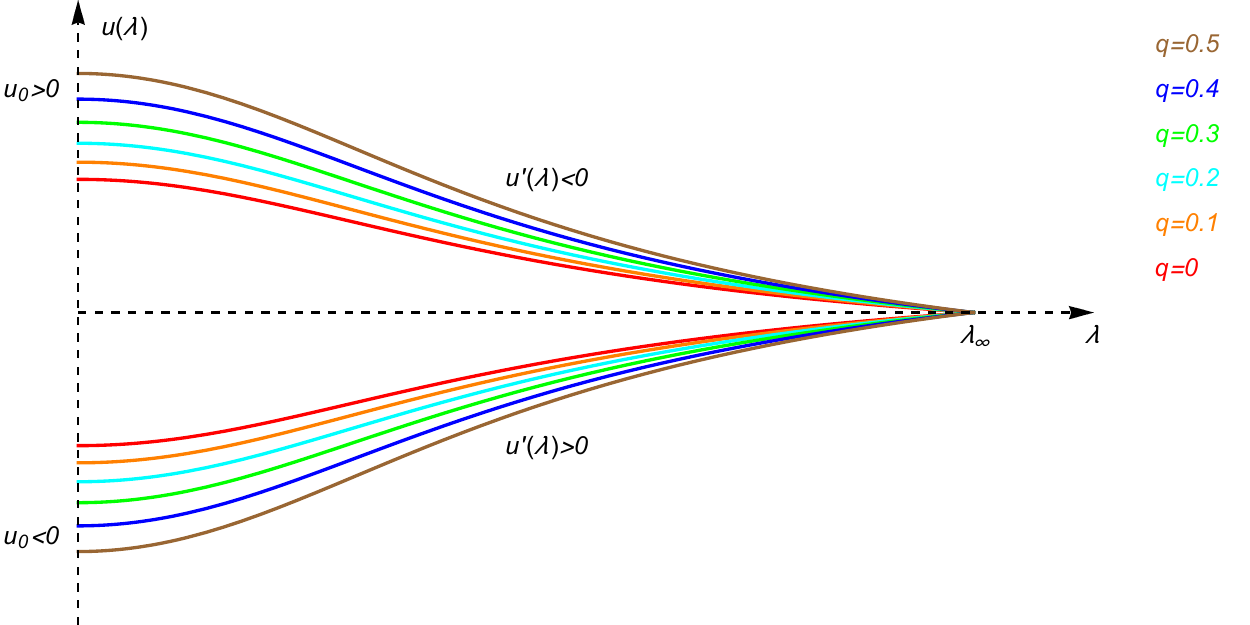}
\par\end{centering}
\caption{\label{fig:3} The minimal embedding function $u\left(\lambda\right)$
of flavor brane with various $q$ and the parameter is set to be $\lambda_{\infty}=5.73r_{KK}$. }
\end{figure}

\begin{figure}
\begin{centering}
\includegraphics[scale=0.38]{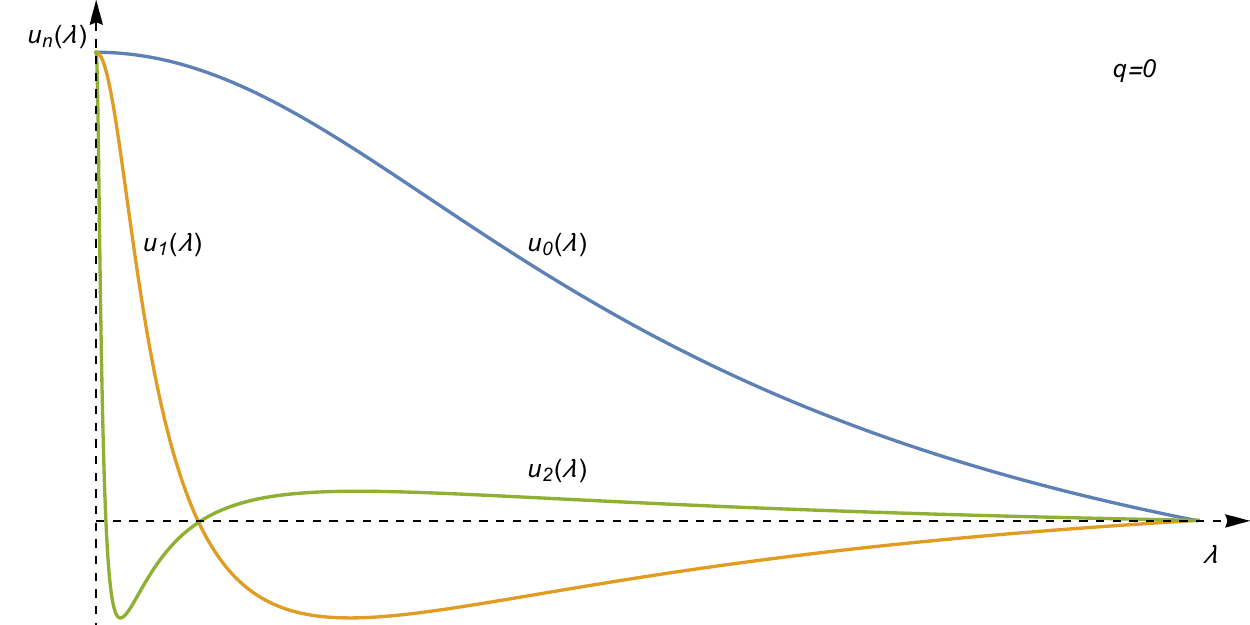}\includegraphics[scale=0.38]{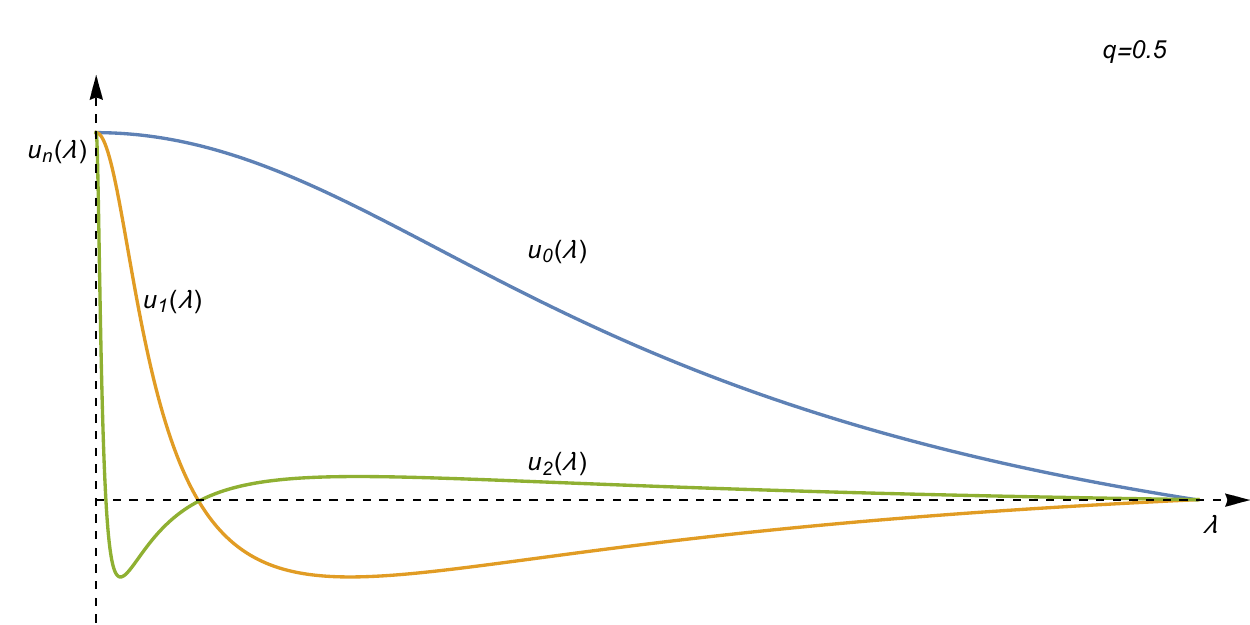}
\par\end{centering}
\begin{centering}
\includegraphics[scale=0.38]{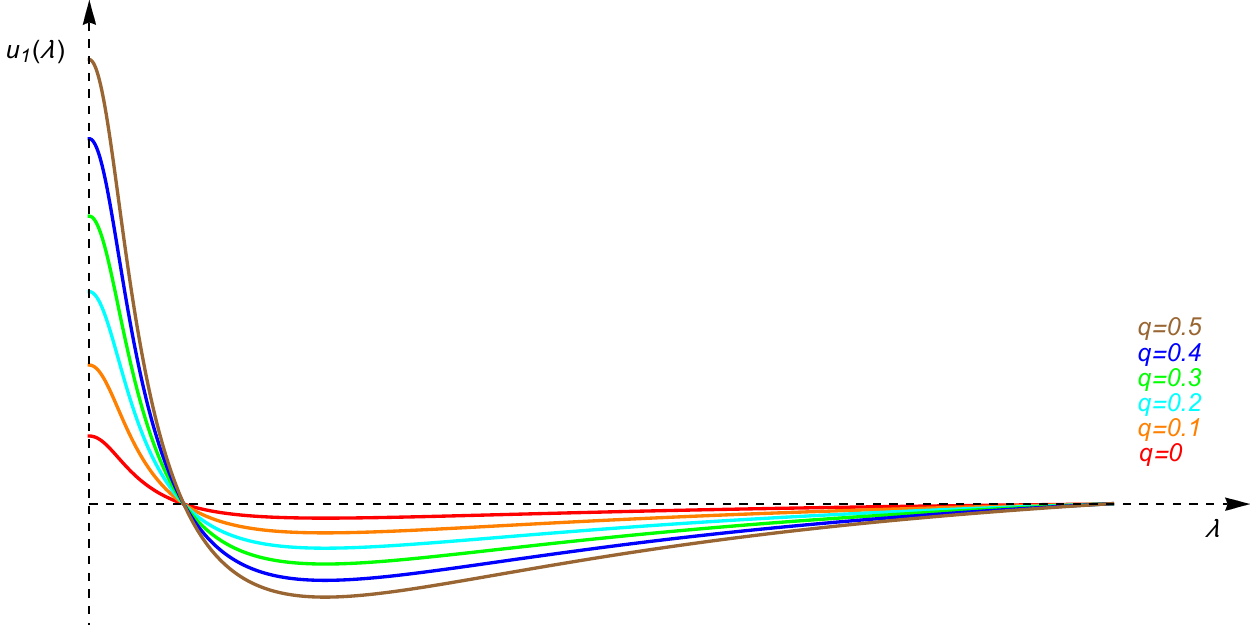}\includegraphics[scale=0.38]{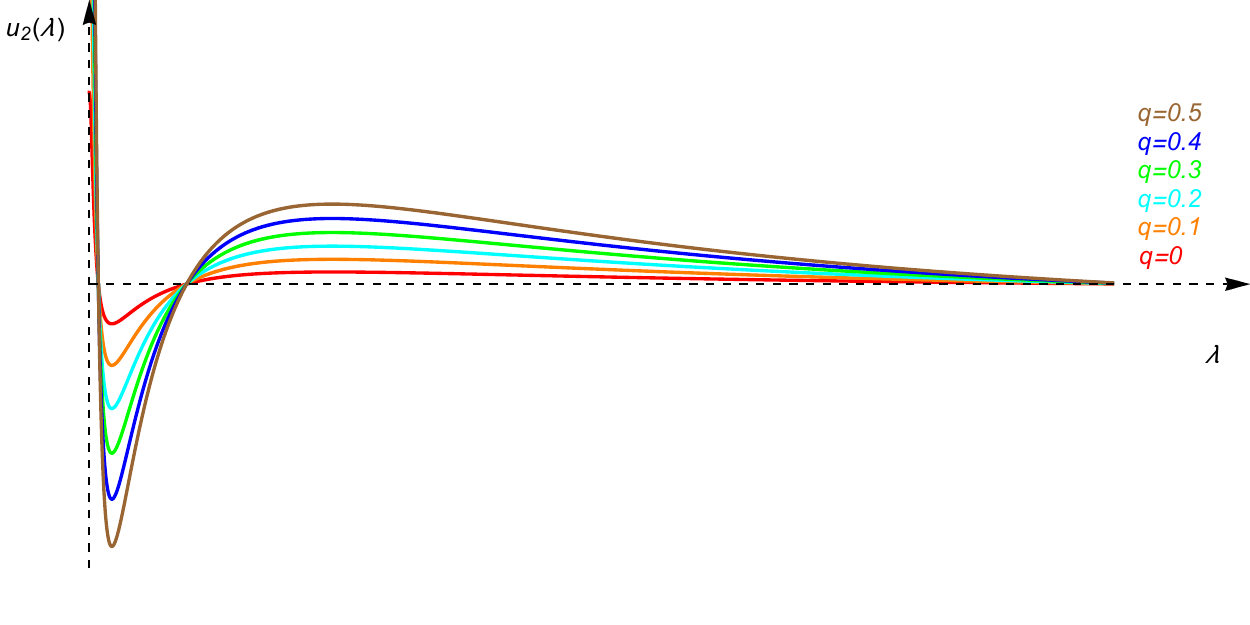}
\par\end{centering}
\caption{\label{fig:4} The upper figures illustrate the relation of $u_{0,1,2}\left(\lambda\right)$
with $q=0,0.5$, the lower figures show the relation of $u_{1,2}$
respected to various $q$.}

\end{figure}

\begin{figure}
\begin{centering}
\includegraphics[scale=0.5]{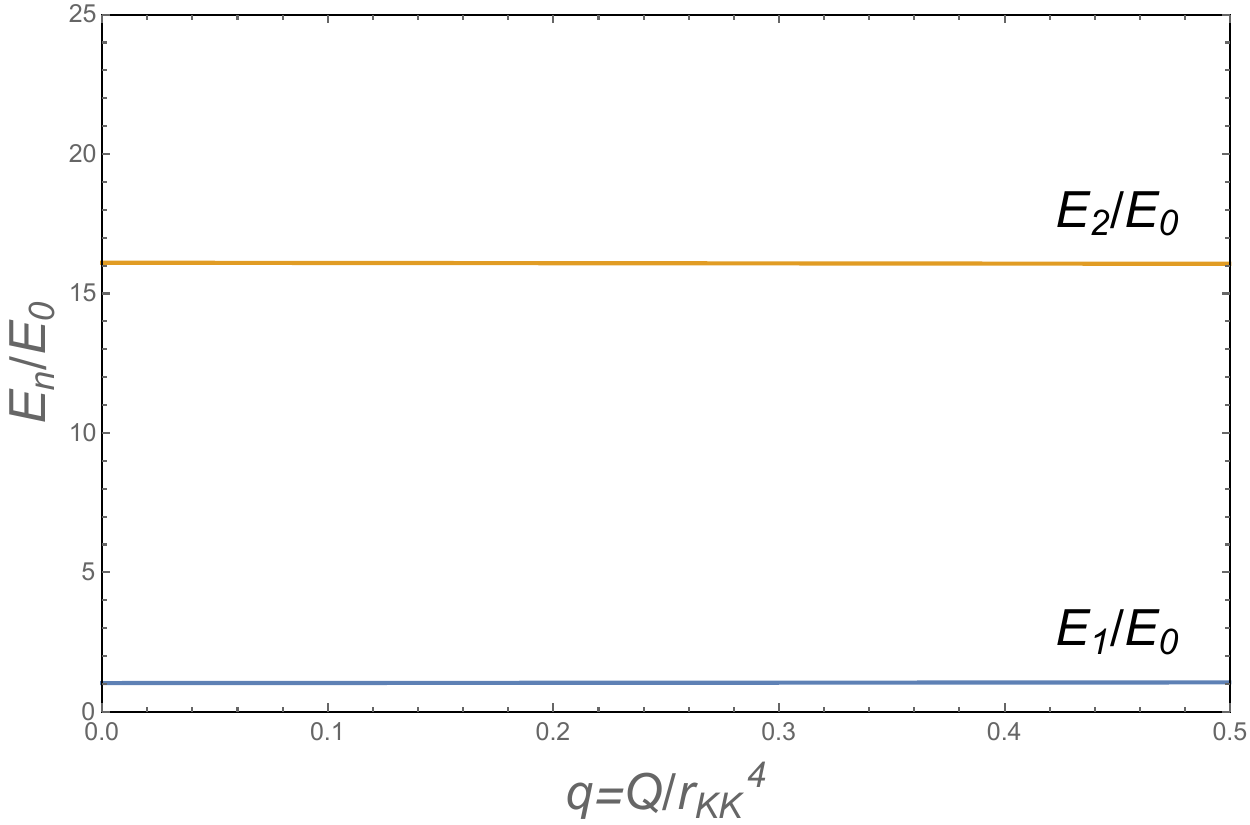}
\par\end{centering}
\caption{\label{fig:5} The ratios of $E_{n}$ and $E_{0}$. $E_{n}$ refers
to the energy of a flavor brane with $n$-nodes embedding function
$u_{n}\left(\lambda\right)$. }

\end{figure}

\begin{figure}
\begin{centering}
\includegraphics[scale=0.5]{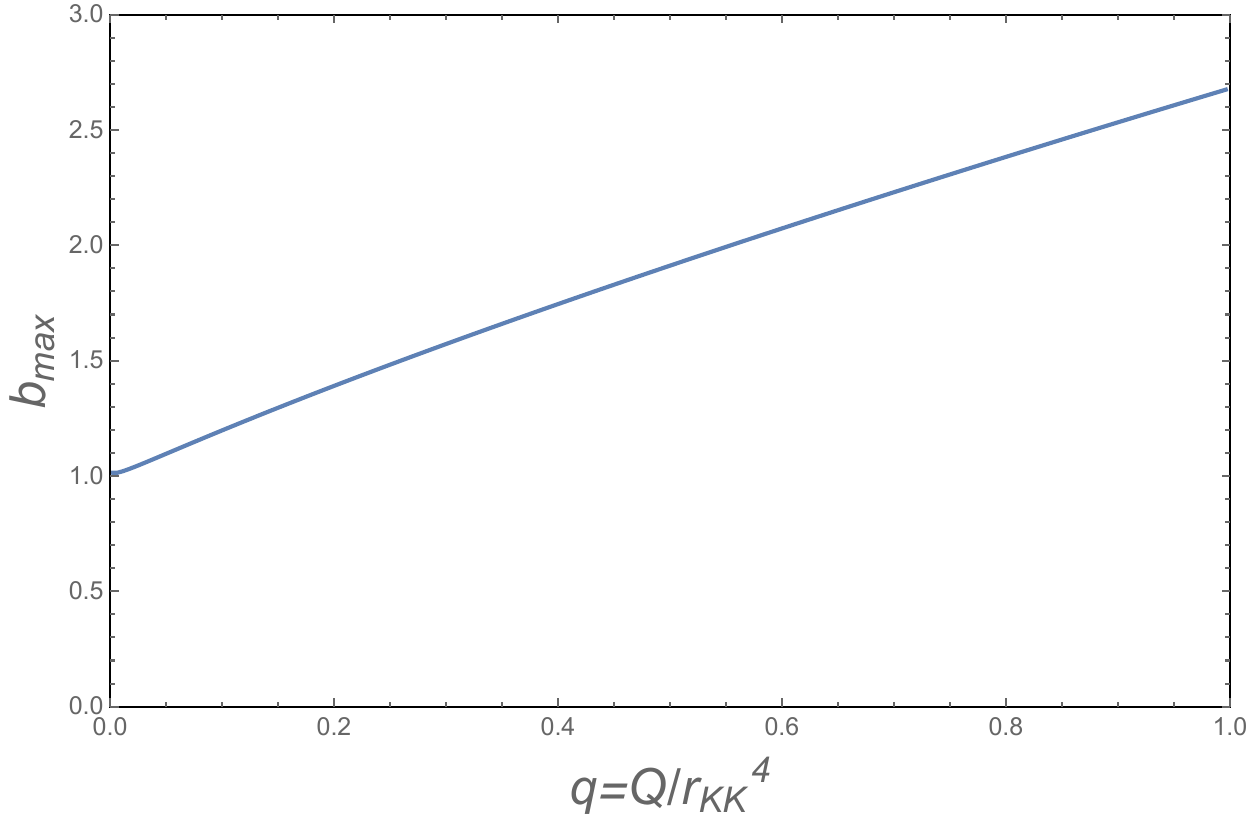}
\par\end{centering}
\caption{\label{fig:6} The relation of $b_{max}$ and $q$. The numerical
value shows $b_{max}\left(0\right)\simeq1.026$ which exactly returns
to the result in \cite{key-15}.}
\end{figure}

In order to find a more general configuration of $u\left(\lambda\right)$,
let us take a look at the asymptotic behaviors of (\ref{eq:17}).
In the region of $\lambda\rightarrow0$, we have $u^{\prime}\left(\lambda\right)\rightarrow0,\rho\rightarrow u,$
so (\ref{eq:17}) leads to a solution as, 

\begin{equation}
u\left(\lambda\right)=\pm u_{0}\mp\left[\frac{4\left(r_{KK}^{4}+u_{0}^{4}\right)}{5u_{0}\left(r_{KK}^{4}+4u_{0}^{4}\right)}-\frac{1}{10}\phi^{\prime}\left(u_{0}\right)\right]\lambda^{2}+\mathcal{O}\left(\lambda^{4}\right).
\end{equation}
Note that $u_{0}>0$ and 

\begin{equation}
\phi^{\prime}\left(u_{0}\right)\equiv\frac{\partial}{\partial\rho}\phi\left(\rho\right)\bigg|_{\rho=u_{0}}=\frac{64Qr_{KK}^{4}u_{0}^{3}}{\left(r_{KK}^{4}-4u_{0}^{4}\right)\left(r_{KK}^{4}+4u_{0}^{4}\right)\left[r_{KK}^{4}+Q\ln\frac{\left(r_{KK}^{4}+4u_{0}^{4}\right)^{2}}{\left(r_{KK}^{4}-4u_{0}^{4}\right)^{2}}\right]}<0,
\end{equation}
due to $u_{0}>\frac{\sqrt{2}}{2}r_{KK}$. The second derivative of
$u\left(\lambda\right)$ takes the opposite sign to $u\left(\lambda\right)\Big|_{\lambda=0}$.
On the other hand, in the region of $\lambda\rightarrow\infty$, we
have $u\left(\lambda\right)\rightarrow u_{\infty},\rho\rightarrow\lambda,$
so the equation (\ref{eq:17}) becomes,

\begin{equation}
\frac{d}{d\lambda}\left(\lambda^{2}u^{\prime}\right)=-2u.
\end{equation}
Accordingly, the asymptotic behavior of $u\left(\lambda\right)$ at
large $\lambda$ takes the general form as,

\begin{equation}
u\left(\lambda\right)=\pm\sqrt{\frac{\mu^{3}}{\lambda}}\sin\left(\frac{\sqrt{7}}{2}\ln\frac{\lambda}{\lambda_{\infty}}\right),
\end{equation}
by imposing the boundary condition $u\left(\lambda_{\infty}\right)=0$
where $\mu,\lambda_{\infty}>0$ are the integration constants. This
solution also has two branches thus it implies the global signs of
$u\left(\lambda\right)$ and $u^{\prime}\left(\lambda\right)$ are
opposite as well. Keeping this in mind, we numerically evaluate the
minimal embedding solution (without any nodes) of (\ref{eq:17}) with
various charge density of D-instantons represented by $q$ and the
results are illustrated as in Figure \ref{fig:3}. 
 Our results show that $q=0$ corresponds to the minimal embedding
among various value of $q$ respected to the solutions with zero node
and this is consistent with that the D3-D-instanton solution describes
the dual theory in an excited background. The equation (\ref{eq:17})
also includes solutions with $n$ nodes and let us denote it as $u_{n}\left(\lambda\right)$,
so the minimal solutions are identified as $u_{0}\left(\lambda,q\right)$
now. We also show the numerical relation of $u_{n=0,1,2}\left(\lambda,q\right)$
for a fixed $q$ and $u_{1,2}\left(\lambda,q\right)$ with various
$q$ as in Figure \ref{fig:4}. 
 The numerical calculation implies that the associated energy of the
embedded flavor brane is a monotonically increased function of the
number of nodes for any $q$, and this is numerically verified as
in Figure \ref{fig:5}. 
 In this sense, the solution of maximal embedding presented in (\ref{eq:20})
can be treated as having infinity nodes. On the other hand, we can
find the total energy of flavor brane is always minimized at $q=0$.
To verify this conclusion quantitatively, we could in particular evaluate
the energy density of a maximal embedded flavor brane since it would
be semi-analytical. Plugging (\ref{eq:20}) into (\ref{eq:15}), we
could obtain the energy density of a maximal embedded flavor brane
as,

\begin{equation}
E_{\mathrm{D7}}^{max}=-\frac{1}{V_{3}}S_{\mathrm{D7}}^{max}=T_{\mathrm{D7}}V_{S^{4}}R^{2}\left[\frac{\lambda_{\infty}}{3}+b_{max}\left(q\right)r_{KK}^{3}\right].\label{eq:26}
\end{equation}
Here $b_{max}\left(q\right)$ is a constant dependent on $q$ which
can be expressed by the combination of generalized hypergeometrical
functions. We plot out the numerical values of $b_{max}\left(q\right)$
as in Figure \ref{fig:6} 
 and it indeed shows $q=0$ corresponds to the the flavor brane of
the lowest energy. In this sense, the vacuum with non-zero $q$ could
be recognized as the metastable vacua of flavors in the presence of
instantons in the dual theory which is in agreement of \cite{key-14,key-15}.

Since we are usually interested in comparing energies with the same
boundary condition between different solutions, the maximal energy
(\ref{eq:26}) should be subtracted as the regularization of the total
energy. Let us denote the onshell action with $n$-nodes embedding
function $u_{n}\left(\lambda\right)$ as $S_{\mathrm{D7}}^{n}$, so
the total energy of the flavor brane is redefined as,

\begin{equation}
E_{f}^{n}\left(q\right)=-\frac{1}{V_{3}}\left(S_{\mathrm{D7}}^{n}-S_{\mathrm{D7}}^{max}\right).\label{eq:27}
\end{equation}
For the minimal embeddings of two parity-related flavor branes, (\ref{eq:27})
could be evaluated as,

\begin{equation}
E_{f}^{0}\left(q\right)\simeq-N_{c}\left(g_{s}N_{c}\right)\left[b_{0}\left(q\right)M_{KK}^{3}+a_{0}\left(q\right)M_{\mu}^{3}\right],\label{eq:28}
\end{equation}
where $b_{0}\left(q\right)$ and $a_{0}\left(q\right)$ depend on
the D-instanton charge $q$. The energy scale $M_{\mu}$ is related
to the length scale $\mu=M_{\mu}R^{2}/2$ which comes from the duality
of holographic radius/energy relation \cite{key-42}. The difference
between the energy density at large $N_{c}$ should be relevant to
the potential barrier of the instanton vacuum in the dual theory.
Besides, the general configuration of $N_{f}$ flavor branes can also
be obtained by (\ref{eq:28}). Let us consider $p$ of $N_{f}$ flavor
branes located in the upper $u,\lambda$ plane while the other $N_{f}-p$
flavor branes located the lower plane with minimal embedding. Since
the energy of each flavor brane is equivalent, the total energy of
these flavor branes should be,

\begin{equation}
E_{f,tot}^{0}=pE_{f}^{0}\left(q\right)+\left(N_{f}-p\right)E_{f}^{0}\left(q\right)=N_{f}E_{f}^{0}\left(q\right).\label{eq:29}
\end{equation}

\subsubsection*{Massive case}

Let us turn to the massive case by considering the inclusion of a
bare mass of the quarks or fermions in the dual theory. The bare mass
in this model can be viewed as a source for the condensate operator
$\left\langle \bar{\psi}\psi\right\rangle $ of fermions. Since the
bare mass of fermions is identified as the spatial separation between
D3- and D7-branes along the transverse direction $u$ in the UV region
and we have seen in the last subsection

\begin{equation}
u\left(\lambda\right)\rightarrow\frac{1}{\sqrt{\lambda}},
\end{equation}
at large $\lambda$, we can set

\begin{equation}
\lim_{\lambda\rightarrow\lambda_{\infty}}\sqrt{\frac{\lambda}{\mu}}u\left(\lambda\right)=2\pi l_{s}^{2}m.
\end{equation}
This is also equivalent to set $u_{\infty}\simeq2\pi l_{s}^{2}m$
while this boundary condition breaks the parity symmetry. So the parity
transformation $u\rightarrow-u$ implies the flavor branes with up/down
embedding have to take positive/negative mass which succeeds the discussion
in the D3/D7 approach consistently. It means the fermion mass leads
to the degeneracy between the upper and lower embedding of the flavor
branes. Then in order to obtain the energy of the flavor brane in
the massive case, we can start from the massless case by performing
a small variation in the boundary condition,

\begin{equation}
\delta u_{\infty}=\sqrt{\frac{\mu}{\lambda_{\infty}}}2\pi l_{s}^{2}\delta m.
\end{equation}
So the associated variation in the onshell action of the flavor brane
becomes,

\begin{align}
\delta S_{\mathrm{D7}} & =\frac{\partial\mathcal{L}}{\partial u^{\prime}}\delta u\bigg|_{\lambda=0}^{\lambda=\lambda_{\infty}}=-T_{\mathrm{D7}}V_{3}V_{S^{4}}R^{2}\left[e^{\phi}\left(r_{KK}^{4}+4\rho^{4}\right)^{3/2}\frac{\lambda^{4}}{8\rho^{8}}\frac{u^{\prime}}{\sqrt{1+u^{\prime2}}}\delta u\right]\bigg|_{\lambda=0}^{\lambda=\lambda_{\infty}},
\end{align}
which leads to

\begin{align}
\delta E_{f} & =T_{\mathrm{D7}}V_{S^{4}}R^{2}\lambda_{\infty}^{2}u^{\prime}\left(\lambda_{\infty}\right)\delta u_{\infty}e^{\phi\left(\lambda_{\infty}\right)}\equiv\mp c\left(q,\lambda_{\infty}\right)\delta m,\nonumber \\
c\left(q,\lambda_{\infty}\right) & =e^{\phi\left(\lambda_{\infty}\right)}\frac{N_{c}\sqrt{g_{s}N_{c}}}{24\pi^{5/2}}M_{\mu}^{2},
\end{align}
where we have used $u^{\prime}\left(0\right)=0$ and the equation
of motion for $u\left(\lambda\right)$. Since the embedding function
can have both signs as in the massless case characterized by $u_{\infty}^{\prime}$,
the positive/negative mass $m$ determines the sign of $u_{\infty}^{\prime}$
as it is preferred. As the massless case, the energy of the flavor
brane can be obtained by using (\ref{eq:27}) which consists of the
massless part $E_{f}^{0}\left(q\right)$ plus a small variation $\delta E_{f}$
as,

\begin{align}
E_{f}^{0}\left(q,m\right) & =E_{f}^{0}\left(q\right)+\delta E_{f}\nonumber \\
 & \simeq E_{f}^{0}\left(q\right)\mp c\left(q,\lambda_{\infty}\right)m.
\end{align}
This result shows that the fermion condensate is negative/positive
for positive/negative mass due to

\begin{equation}
\left\langle \bar{\psi}\psi\right\rangle =\frac{dE_{f}^{0}\left(q,m\right)}{dm}=\mp c\left(q,\lambda_{\infty}\right)\mathrm{sign}\left(m\right).
\end{equation}
And it would be slightly modified by the presence of the D-instantons
in the quadratic order of the fermion mass.

To close this subsection, let us evaluate the total energy of $N_{f}$
flavor branes with a bare mass. As before, we consider the configuration
that $p$ of $N_{f}$ branes wraps the upper half-five-sphere separated
from the other $N_{f}-p$ branes wrapping the lower half-five-sphere
with a common mass $m$. So the total energy is a collection of upper
and lower branes which is,

\begin{align}
E_{f,tot}^{0}\left(q,m\right) & =p\left[E_{f}^{0}\left(q\right)-c\left(q,\lambda_{\infty}\right)m\right]+\left(N_{f}-p\right)\left[E_{f}^{0}\left(q\right)+c\left(q,\lambda_{\infty}\right)m\right]\nonumber \\
 & =N_{f}E_{f}^{0}\left(q\right)+\left(N_{f}-2p\right)c\left(q,\lambda_{\infty}\right)m.\label{eq:37}
\end{align}
It would be obvious that, for any $q$, the minimal energy occurs
at $p=N_{f}$ for $m>0$ and $p=0$ for $m<0$. And (\ref{eq:37})
reduces to (\ref{eq:29}) if $m\rightarrow0$ so that the degeneracy
of upper and lower embeddings is regained.

\subsection{Embedding of the CS D7-brane}

Since the R-R flux $C_{0}$ is non-vanished in our D3-D(-1) background,
there should also be a magnetic source for $C_{0}$. And the source
could be provided by $n_{b}$ CS D7-branes as probes coupled to $C_{0}$
magnetically. The configuration of the CS brane is illustrated in
Table \ref{tab:1}. Since the number of CS branes should be an integer,
the CS level is automatically quantized in holography. 

\begin{figure}
\begin{centering}
\includegraphics[scale=0.38]{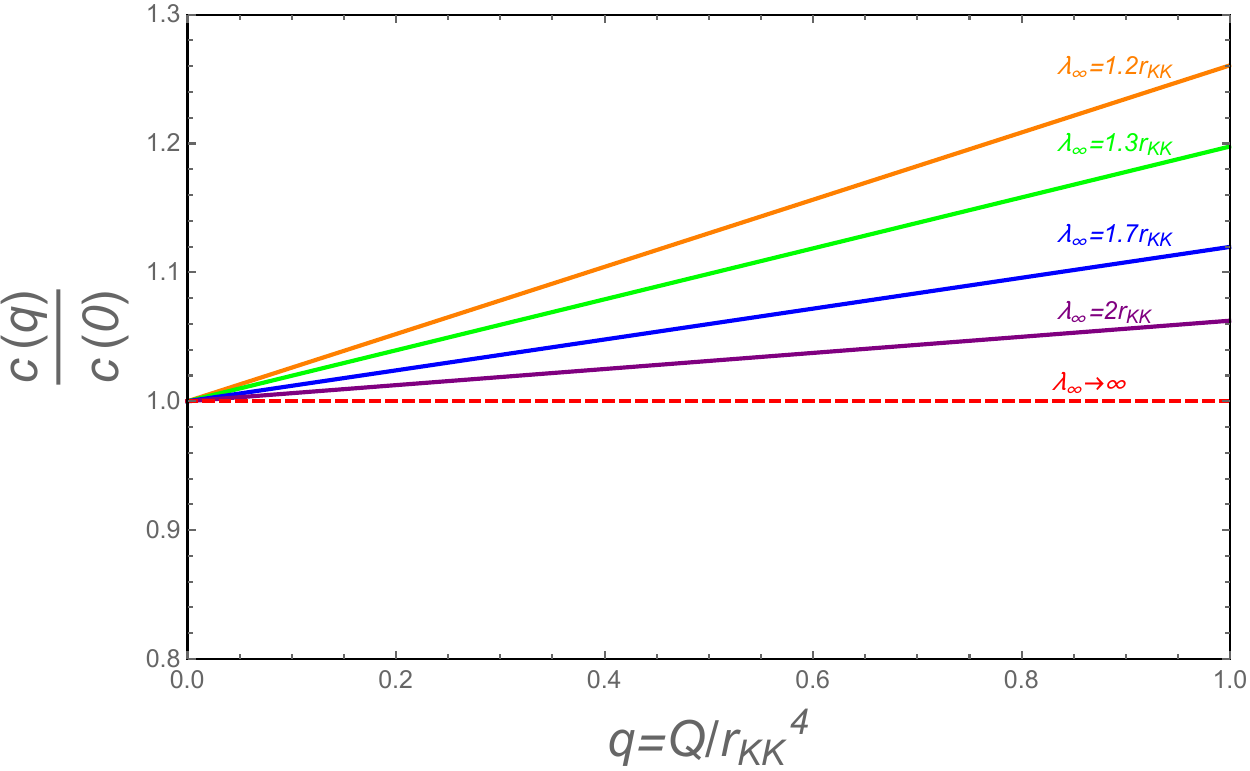}\includegraphics[scale=0.38]{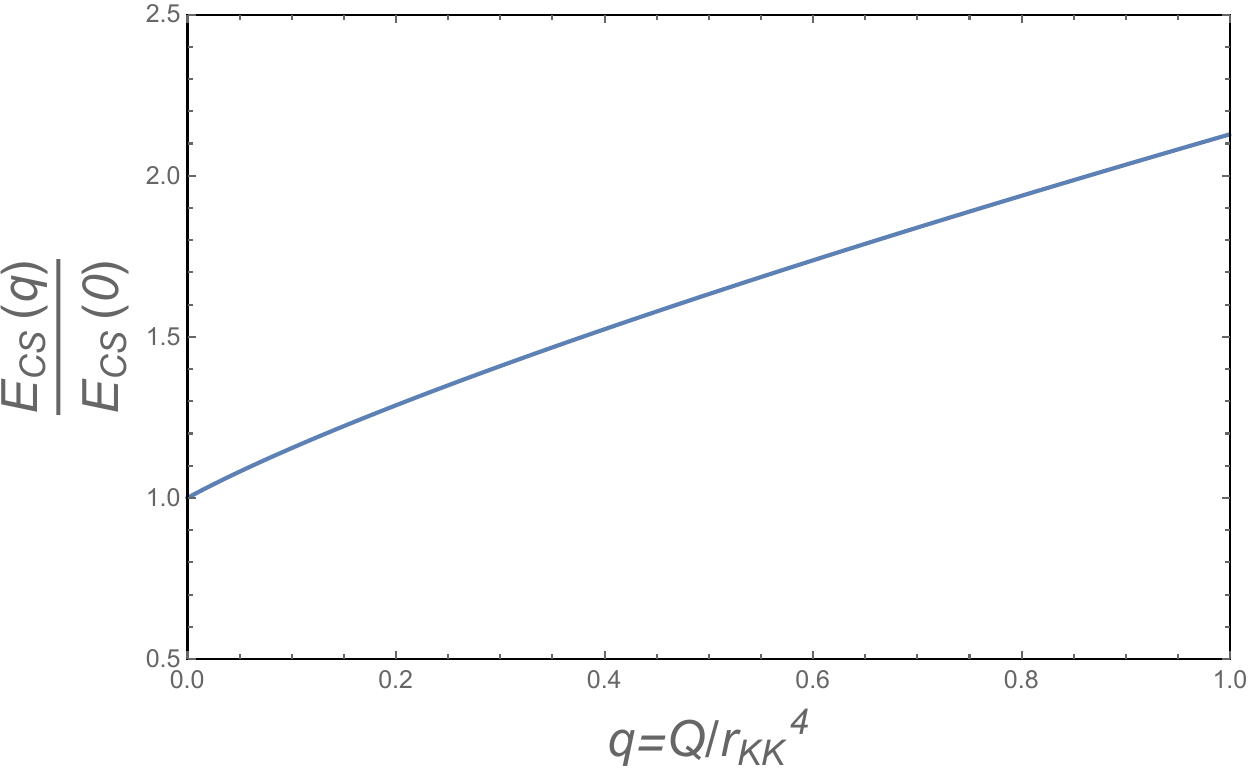}
\par\end{centering}
\caption{\label{fig:7} The relation of $c\left(q,\lambda_{\infty}\right)$
(left), $G\left(q\right)=E_{\mathrm{CS}}\left(q\right)/E_{\mathrm{CS}}\left(0\right)$
(right) and $q$. For $q=0$, it returns to the D3-brane approach
without any D-instantons.}
\end{figure}

In the D3-brane approach, the CS brane can be set located at $r=r_{KK}$
in order to minimize their energy density, however this does not work
in the presence of D-instantons because the energy density of a single
CS brane is evaluated as, 

\begin{align}
S_{\mathrm{D}7}^{\mathrm{CS}} & =-T_{\mathrm{D}7}\int d^{8}xe^{-\phi}\sqrt{-g_{\mathrm{D7}}}-\mu_{7}\int C_{8}=-T_{\mathrm{D}7}V_{3}V_{S^{5}}R^{2}r^{3}e^{\phi}-N_{\mathrm{D}},
\end{align}
where $C_{8}$ is the dual form of $C_{0}$ defined as $dC_{8}=^{\star}dC_{0}$.
This action is divergent at $r=r_{KK}$ which leads to an IR divergence
in the dual field. While this is not important when we are interested
in comparing the difference of the energy, the position of the CS
brane would be less clear. To figure out this problem we require that
our discussion should reduce to the D3-brane approach if $q\rightarrow0$.
In this sense, we assume that the location of the CS brane $r=r_{KK}$
is shifted by the presence of D-instantons which becomes $r=r_{Q}>r_{KK}$.
And we furthermore treat the solution (\ref{eq:20}) describing a
CS D7-brane embedding at $r=r_{Q}$ according to the embedding equation
(\ref{eq:17}), so that

\begin{equation}
r_{Q}=\frac{1}{\sqrt{2}}\left(k+\frac{1}{k}\right)^{1/2}r_{KK}.
\end{equation}
Hence for a fixed $q$, the maximal embedding of a flavor D7-brane
can be identified as an embedding function of CS D7-brane as it is
done in the D3-brane approach. The positive and negative level of
the CS brane corresponds to the orientation of counterclockwise and
clockwise respectively in the $u,\lambda$ plane. Therefore the total
energy of a CS brane can be evaluated as,

\begin{equation}
E_{\mathrm{CS}}\left(q\right)=-\frac{S_{\mathrm{D7}}^{\mathrm{CS}}}{V_{3}}=\mu_{7}V_{S^{5}}R^{2}r_{Q}^{3}e^{\phi\left(r_{Q}\right)}=\frac{g_{s}N_{c}^{2}M_{KK}^{3}}{64\pi^{2}}\left[\frac{1}{2\sqrt{2}}\left(k+\frac{1}{k}\right)^{3/2}e^{\phi\left(r_{Q}\right)}+\frac{1}{2}q\right],\label{eq:40}
\end{equation}
which indeed reduces to the D3-brane approach when $q\rightarrow0$.
We plot out $G\left(q\right)=E_{\mathrm{CS}}\left(q\right)/E_{\mathrm{CS}}\left(0\right)$
and $c\left(q,\lambda_{\infty}\right)/c\left(0,\lambda_{\infty}\right)$
as a function of $q$ in Figure \ref{fig:7}. 
 The numerical result also illustrates $q=0$ corresponds to the CS
brane with minimal energy and the limit of $r_{Q}\rightarrow r_{KK}$.

At low energy, the CS brane reduces to a 3d $U\left(n_{b}\right)$
gauge theory and leads to a Chern-Simons action at level $-N$ due
to the Wess-Zumino term of the D-brane action which is,

\begin{equation}
S_{C_{4}}=\frac{1}{2\left(2\pi\right)^{5}l_{s}^{4}}\int_{\mathrm{D7}}C_{4}\land\mathrm{Tr}\left(F\wedge F\right)=-\frac{1}{2\left(2\pi\right)^{5}l_{s}^{4}}\int_{S^{5}}F_{5}\int_{\mathbb{R}^{1,2}}\mathrm{Tr}\omega_{3}=-\frac{N_{c}}{4\pi}\int_{\mathbb{R}^{1,2}}\mathrm{Tr}\omega_{3}.
\end{equation}
All excitations on the CS branes will decouple at very low energy
scale except this CS term, so we can obtain the level/rank duality
$SU\left(N\right)_{n_{b}}\leftrightarrow U\left(\left|n_{b}\right|\right)_{N}$
through this holographic system as the QFT expectations.

\section{Vacuum structure of the dual theory}

In this section, let us analyze the vacuum structure in the dual theory
at large-$N_{c}$ expansion. Since the vacuum of the dual theory in
general may include both flavor and CS branes, we are going to take
into account the configuration with two kinds of the D7-branes. 

\begin{figure}
\begin{centering}
\includegraphics[scale=0.35]{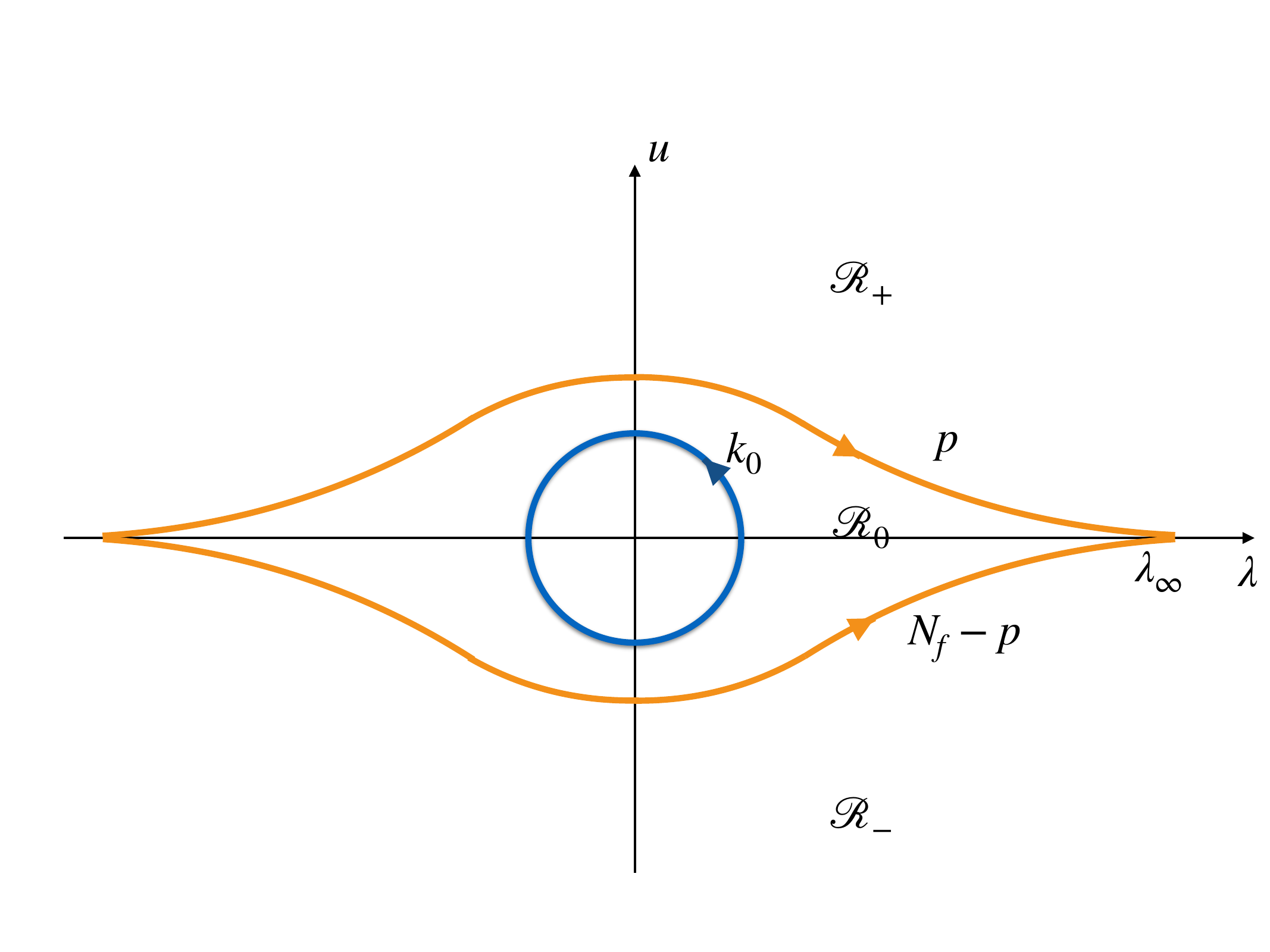}
\par\end{centering}
\caption{\label{fig:8} The configuration of flavor and CS branes on $u,\lambda$
plane. The flavor branes are represented by the orange line and the
CS brane are represented by the blue line. $\mathcal{R}_{+},\mathcal{R}_{-},\mathcal{R}_{0}$
respectively refers to the regions above, between and below the flavor
branes. }
\end{figure}

\begin{figure}
\begin{centering}
\includegraphics[scale=0.5]{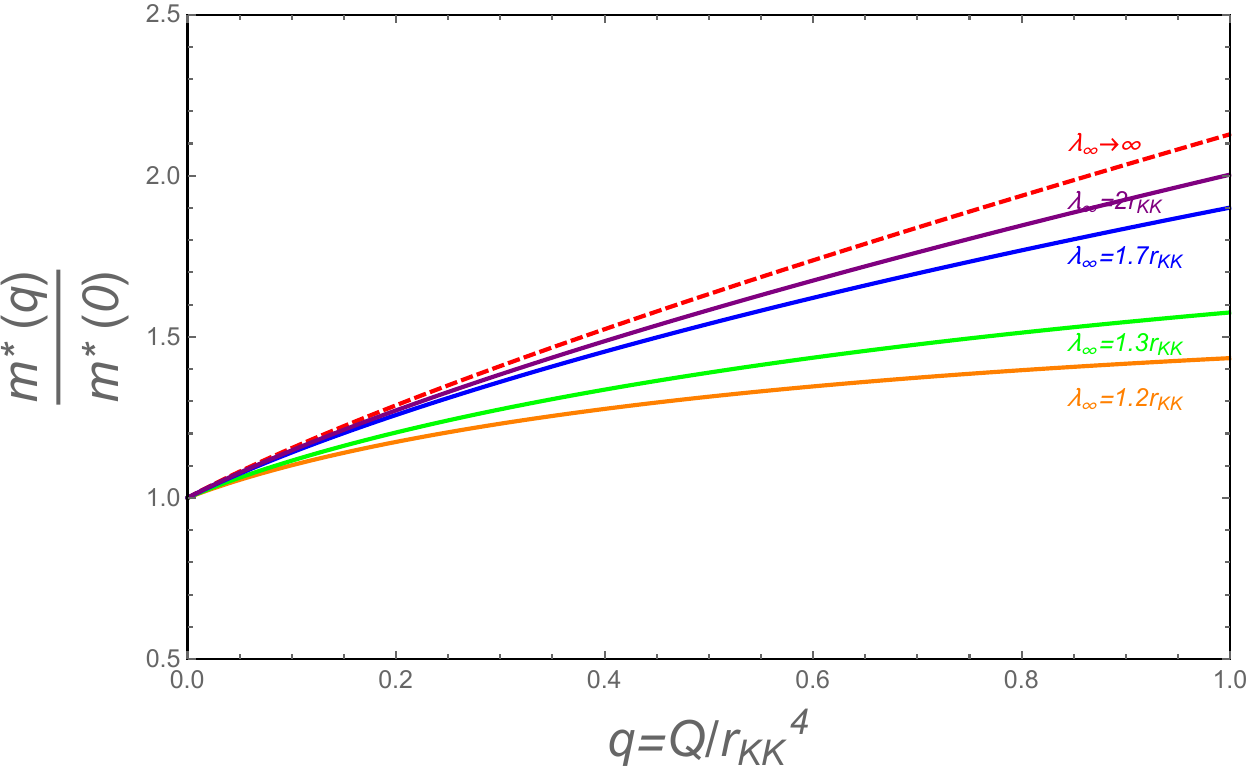}
\par\end{centering}
\caption{\label{fig:9} Relation of $m^{*}\left(q\right)$ and $q$. The critical
mass for the phase transition is increased by the presence of the
D-instantons. }
\end{figure}

The effective CS level in the dual theory is required as,

\begin{equation}
i\int_{S^{1}}F_{1}\big|_{r\rightarrow\infty}=k_{eff}.
\end{equation}
However in order to define the CS level $\kappa=k_{b}-N_{f}/2$, the
CS level must depend on $p$. To find the result, we can straightforwardly
count the number of the orientation in the $u,\lambda$ plane. Defining
the number of D7-branes with counterclockwise/clockwise orientation
is positive/negative, let us consider the configuration of that, in
the $u,\lambda$ plane $k_{0}$ counterclockwise CS branes live in
$\mathcal{R}_{0}$, $p$ flavor branes live in $\mathcal{R}_{+}$
and $N_{f}-p$ flavor branes live in $\mathcal{R}_{-}$ where $\mathcal{R}_{+},\mathcal{R}_{-},\mathcal{R}_{0}$
respectively refers to the regions of the $u,\lambda$ plane which
are above, between and below the flavor branes as it is illustrated
in Figure \ref{fig:8}. 
 We only consider the minimal embedding of the flavor branes since
the concern is the vacuum structure. Requiring $\kappa=k_{eff}$ at
the UV boundary, we have \cite{key-15},

\begin{equation}
k_{eff}=\begin{cases}
\kappa-\frac{N_{f}}{2}, & \mathrm{in}\ \mathcal{R}_{+},\\
\kappa+p-\frac{N_{f}}{2}, & \mathrm{in}\ \mathcal{R}_{0},\\
\kappa+\frac{N_{f}}{2}, & \mathrm{in}\ \mathcal{R}_{-},
\end{cases}
\end{equation}
and $k_{0}=\kappa+p-N_{f}/2$ which is what we desire in the dual
field theory. The interpretation of such D-brane configuration at
low energy is that, the flavor symmetry $U\left(N_{f}\right)$ is
broken spontaneously to $U\left(p\right)\times U\left(N_{f}-p\right)$
which creates $2p\left(N_{f}-p\right)$ Goldstone bosons and their
target space is Grassmann, 

\begin{equation}
\mathrm{Gr}\left(p,N_{f}\right)=\frac{U\left(N_{f}\right)}{U\left(p\right)\times U\left(N_{f}-p\right)}.
\end{equation}
The CS branes leads to a level/rank duality of $U\left(\left|\kappa+p-N_{f}/2\right|\right)_{N}\leftrightarrow SU\left(N\right)_{k+p-N_{f}/2}$.
So the low-energy dynamics of a $p$ sector would have the symmetry,

\begin{equation}
\mathrm{Gr}\left(p,N_{f}\right)\times SU\left(N\right)_{k+p-N_{f}/2},
\end{equation}
in which the $N_{f}+1$ sectors describe the vacuum of the dual theory
holographically. To analyze the phase structure of the vacuum, the
minimal value of the (free) energy is necessary. Since the total energy
of the $p$ sector consists of flavor plus the CS part and the flavor
energy has be obtained in (\ref{eq:37}), we need to include the energy
of the CS brane which is the number of CS branes times the energy
density $E_{\mathrm{CS}}\left(q\right)$ given in (\ref{eq:40}).
Therefore the total free energy density is collected as ($\kappa\geq0,0\leq p\leq N_{f}$)
\cite{key-15},

\begin{equation}
E\left(p,q\right)=N_{f}E_{f}^{0}\left(q\right)+\left(N_{f}-2p\right)c\left(q,\lambda_{\infty}\right)m+\left|\kappa+p-N_{f}/2\right|E_{\mathrm{CS}}.\label{eq:46}
\end{equation}
Minimize (\ref{eq:46}) then compare the free energy, the associated
free energy density is obtained as, for $\kappa>N_{f}/2$,

\begin{align}
E_{vac} & =\begin{cases}
\left(\kappa-N_{f}/2\right)E_{\mathrm{CS}}, & m<m^{*},\\
\left(\kappa+N_{f}/2\right)E_{\mathrm{CS}}-2N_{f}cm, & m>m^{*},
\end{cases}\begin{array}{c}
SU\left(N\right)_{k-N_{f}/2},\\
SU\left(N\right)_{k+N_{f}/2},
\end{array}\label{eq:47}
\end{align}
where $k\pm N_{f}/2$ refers to the corresponding topological phase
in the dual theory. And for $\kappa<N_{f}/2$, the minimized free
energy density and associated topological phase are collected as,

\begin{align}
E_{vac} & =\begin{cases}
\left(N_{f}/2-\kappa\right)E_{\mathrm{CS}}, & m<-m^{*},\\
2\left(\kappa-N_{f}/2\right)c\left(q,\lambda_{\infty}\right)m, & -m^{*}<m<m^{*},\\
\left(N_{f}/2+\kappa\right)E_{\mathrm{CS}}-2N_{f}c\left(q,\lambda_{\infty}\right)m, & m>m^{*},
\end{cases}\begin{array}{c}
SU\left(N\right)_{k-N_{f}/2},\\
\mathrm{Gr}\left(p,N_{f}\right),\\
SU\left(N\right)_{k+N_{f}/2},
\end{array}\label{eq:48}
\end{align}
where the critical mass $m^{*}$ is defined as,

\begin{equation}
m^{*}=\frac{E_{\mathrm{CS}}\left(q\right)}{2c\left(q,\lambda_{\infty}\right)}.
\end{equation}
Since $c\left(q,\lambda_{\infty}\right)$ never goes to zero, the
derivative with respect to $m$ both in (\ref{eq:47}) and (\ref{eq:48})
is discontinuous which means there remains to be a first order phase
transition at $m=\pm m^{*}$ in the presence of the D-instantons.
And the vacua would be degenerate at the critical point. However in
our holographic approach, the mass $m^{*}$ additionally depends on
the charge density $q$ of the D-instantons. So we numerically evaluate
$m^{*}$ as a function of $q$ as in Figure \ref{fig:9}. 
 According to the numerical calculation, while the order parameter
$c\left(q,\lambda_{\infty}\right)$ in the UV limit $\lambda_{\infty}\rightarrow0$
is almost unchanged, the critical mass $m^{*}$ is increased by the
presence of the D-instantons which shifts the phase transition point
for $\kappa>N_{f}/2$ thus enhance the phase of $\mathrm{Gr}\left(p,N_{f}\right)$
for $\kappa<N_{f}/2$. Accordingly, the D3-D(-1) approach implies
the phase transition point is also determined by the D-instanton charge.
This could be interpreted as the topological effect in the dual theory
which is similar as the topological contribution to the mass in the
presence of the CS term \cite{key-28}. 

\section{Entanglement entropy and confinement }

Since the entanglement entropy is expected to be a tool to characterize
the confinement/deconfinement phases of the dual theory \cite{key-37,key-38,key-39,key-40},
in this section we will compute the quantum entanglement entropy between
two physically disjoint spatial regions in the bulk, then compare
the results with the analysis of the free energy. 

Before the holographic calculation, we first take into account the
simplest geometry: region $A$ is the product of $\mathbb{R}^{2}\times I_{l}$
where $I_{l}$ is a line interval of length $l$ and region $B$ is
the complement of $A$. According to the AdS/CFT dictionary, the quantum
entanglement entropy between region $A$ and $B$ relates to the surface
$\gamma$ in bulk whose boundary coincides with the boundary of $A$.
Supposing we are discussing the correspondence of AdS\textsubscript{d+2}/CFT\textsubscript{d+1},
the classical area of surface $\gamma$ is given as,

\begin{equation}
S_{\gamma}=\frac{1}{4G_{N}^{d+2}}\int_{\gamma}d^{d}x\sqrt{g_{\mathrm{ind}}},\label{eq:50}
\end{equation}
where $G_{N}^{d+2}$ is the $d+2$ dimensional Newton constant and
$g_{\mathrm{ind}}$ refers to the induced metric on $\gamma$. Notice
$\gamma$ has to be spatial like to represent the entanglement entropy
at a fixed time. The (\ref{eq:50}) can also be generalized into non-conformal
situations. For example, in 10d geometry of D-branes, (\ref{eq:50})
could be naturally modified as,

\begin{equation}
S_{A}=\frac{1}{4G_{N}^{10}}\int_{\gamma}d^{8}x\sqrt{g_{\mathrm{ind}}}.\label{eq:51}
\end{equation}
We will use (\ref{eq:51}) to evaluate the the quantum entanglement
entropy in our holographic model. 

\begin{figure}
\begin{centering}
\includegraphics[scale=0.38]{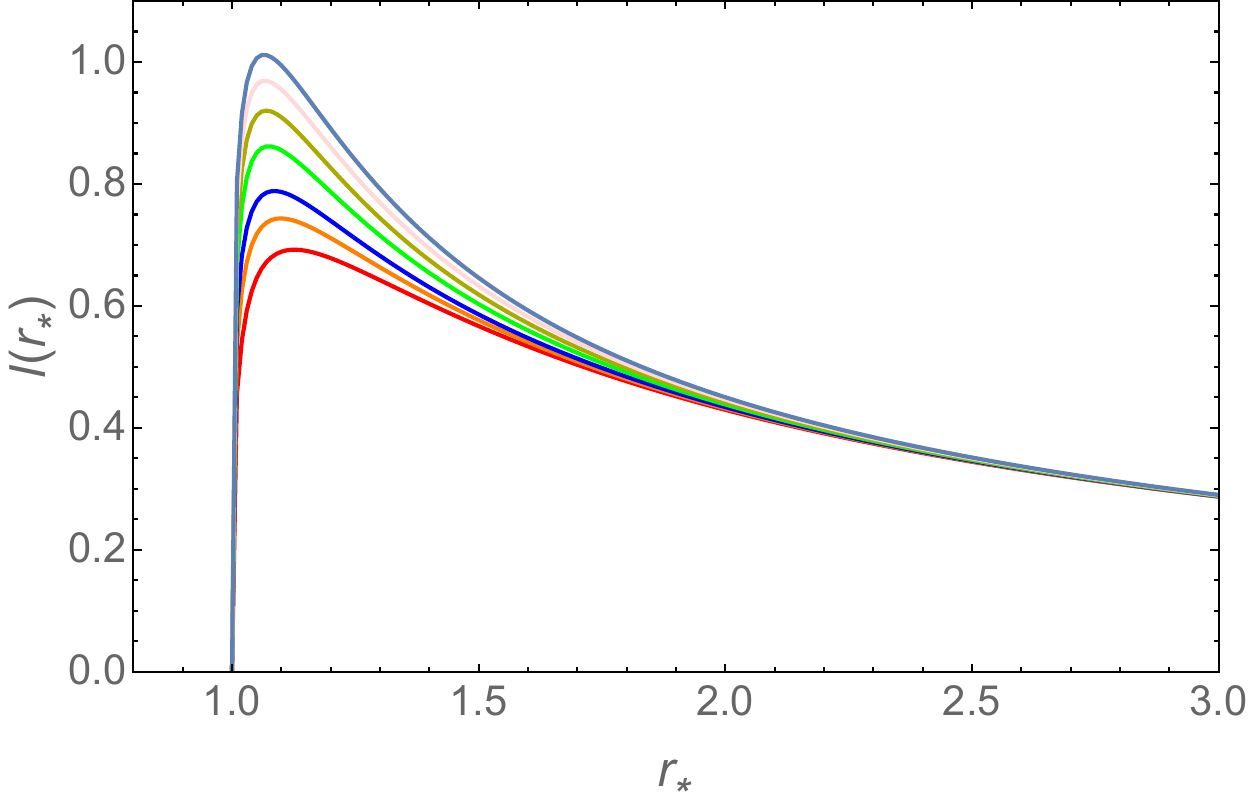}\includegraphics[scale=0.38]{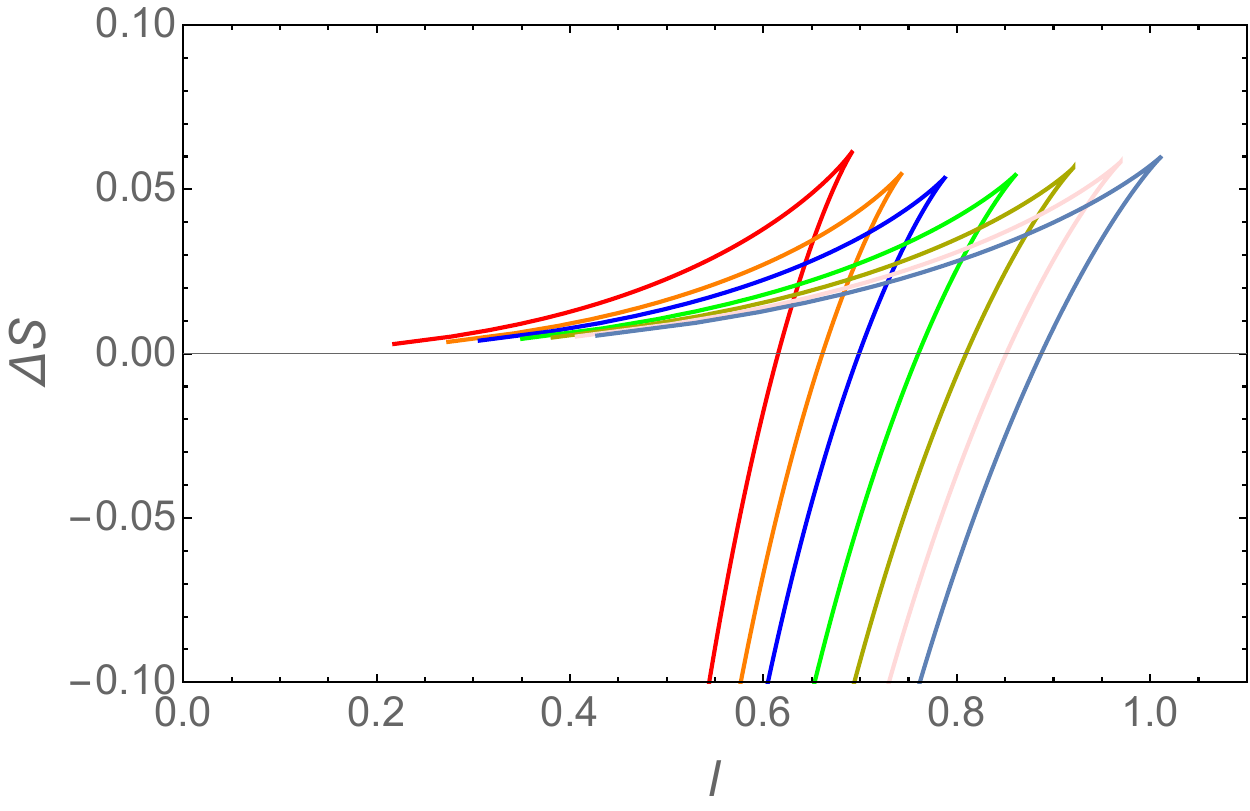}
\par\end{centering}
\caption{\label{fig:10} Left: Relation of $l$ and $r_{*}$ for $q=0,0.5,1,2,3,4,5$
(lower to upper), $l_{\mathrm{max}}$ is increased by $q$. Right:
Relation of $\Delta S$ and $l$ for $q=0,0.5,1,2,3,4,5$ (left to
right) as a typical swallow-tail behavior. The critical length $l_{c}$
is also increased by $q$.}

\end{figure}

The most convenient way to begin the calculation is to write the 10d
metric as,

\begin{equation}
ds^{2}=\alpha\left(r\right)\left[\beta\left(r\right)dr^{2}+\eta_{\mu\nu}dx^{\mu}dx^{\nu}\right]+g_{mn}dy^{m}dy^{n},\label{eq:52}
\end{equation}
where $\mu=0,1,...d$, $m=d+2,...9$ parametrize $\mathbb{R}^{d+1}$
and $8-d$ internal directions respectively, $r$ refers to the holographic
radial coordinate. Using (\ref{eq:51}) with formula (\ref{eq:52}),
the minimized action is given as,

\begin{equation}
S_{A}=\frac{V_{d-1}}{2G_{N}^{10}}\int_{r_{*}}^{r_{\infty}}dr\frac{\sqrt{\beta\left(r\right)}H\left(r\right)}{\sqrt{H\left(r\right)-H\left(r_{*}\right)}},\label{eq:53}
\end{equation}
where

\begin{align}
l\left(r_{*}\right) & =2\sqrt{H\left(r_{*}\right)}\int_{r_{*}}^{\infty}dr\frac{\sqrt{\beta\left(r\right)}}{\sqrt{H\left(r\right)-H\left(r_{*}\right)}},\nonumber \\
H\left(r\right) & =e^{-4\phi}V_{\mathrm{int}}^{2}\alpha^{d}\left(r\right),\nonumber \\
V_{\mathrm{int}} & =\int\prod_{m=1}^{8-d}dy^{m}\sqrt{\det g}.
\end{align}
The minimal surface has distinct features for small and large $l$
according to the definition of region $A$ and $B$. The minimal surface
extends into the bulk up to the radial position $r_{*}>r_{KK}$ as
a connected surface for small $l$ while the minimal surface becomes
two disconnected pieces and extends in the bulk all the way up to
$r_{KK}$ for large $l$. In order to characterize the phase transition,
we need to compare the entanglement entropy of connected with disconnected
configuration of the minimal surface. While the entanglement entropy
itself may be divergent, its difference $\Delta S$ could be finite
which according to (\ref{eq:53}) could be written as,

\begin{align}
\Delta S & \equiv\frac{2G_{N}^{10}}{V_{d-1}}\left(S_{A}^{\mathrm{conf}}-S_{A}^{\mathrm{dic}}\right)\nonumber \\
 & =\int_{r_{*}}^{\infty}dr\sqrt{\beta\left(r\right)H\left(r\right)}\left\{ \left[1-\frac{H\left(r_{*}\right)}{H\left(r\right)}\right]^{-1/2}-1\right\} -\int_{r_{KK}}^{r_{*}}dr\sqrt{\beta\left(r\right)H\left(r\right)}.\label{eq:55}
\end{align}
Plugging (\ref{eq:5}) into (\ref{eq:52}) - (\ref{eq:55}), we can
numerically calculate the relation of $\Delta S$ and $l$, $l$ and
$r_{*}$ as illustrated in Figure \ref{fig:10}. 
 As we can see, the critical length $l_{c}$ ($\Delta S=0$) is increased
by the density of the D-instanton charge denoted by $q$, which implies
there would be a first order phase transition at scale $l_{c}$ and
it is enhanced in the presence of the D-instantons. Since the entanglement
entropy takes order of $\mathcal{O}\left(N_{c}^{0}\right),\mathcal{O}\left(N_{c}^{2}\right)$
respectively for $l>l_{c}$ and $l<l_{c}$, this phase transition
may probably relates to the deconfinement phase transition in a QCD-like
theory. So it would be, on the other hand, very interesting to evaluate
the critical temperature of the deconfinement phase transition to
examine whether it is consistent with the analyses of the entanglement
entropy in holography.

To obtain the critical temperature of the deconfinement in the dual
theory, we should compute the associated free energy $F$ in holography
which is the summary of the Euclidean version of the onshell action
(\ref{eq:1}) denoted as $S_{\mathrm{IIB}}^{E}$, Gibbons-Hawking
term $S_{\mathrm{GH}}$ and holographic counterterm $S_{\mathrm{ct}}^{\mathrm{bulk}}$
in bulk, since the deconfinement phase transition is suggested to
be identified as the Hawking-Page transition in the bulk \cite{key-33,key-34,key-35,key-36}.
In Einstein frame, they are given as \cite{key-43,key-44},

\begin{align}
S_{\mathrm{bulk}} & =S_{\mathrm{IIB}}^{E}+S_{\mathrm{GH}}+S_{\mathrm{ct}}^{\mathrm{bulk}},\nonumber \\
S_{\mathrm{IIB}}^{E} & =-\frac{1}{2\kappa_{10}^{2}}\int d^{10}x\sqrt{g}\left[\mathcal{R}-\frac{1}{2}\partial\Phi\cdot\partial\Phi-\frac{1}{2}e^{2\Phi}\left|F_{1}\right|^{2}-\frac{1}{2}\left|F_{5}\right|^{2}\right],\nonumber \\
S_{\mathrm{GH}} & =-\frac{1}{\kappa_{10}^{2}}\int d^{9}xe^{-2\Phi_{0}}\sqrt{h}\left(K-K_{0}\right),\nonumber \\
S_{\mathrm{ct}}^{\mathrm{bulk}} & =-\frac{1}{\kappa_{10}^{2}}\int d^{9}xe^{-2\Phi_{0}}\frac{5}{R}\sqrt{h},\label{eq:56}
\end{align}
where $h$ is the determinant of the boundary metric i.e. the slice
of the 10d metric (\ref{eq:5}) in Einstein frame at fixed $r=r_{\infty}$
with $r_{\infty}\rightarrow\infty$. $K$ is the trace of the extrinsic
curvature at the boundary and $K_{0}$ arises from the standard transformation
of the gravity action from Einstein to string frame and they are given
as,

\begin{equation}
K=-\frac{1}{\sqrt{g}}\partial_{r}\sqrt{h}\big|_{r\rightarrow\infty},\ K_{0}=\frac{9}{4}\sqrt{g^{rr}}\partial_{r}\phi\big|_{r\rightarrow\infty}.
\end{equation}

Then to include the contribution of the flavors and CS level, we additionally
need to evaluate the Euclidean onshell action of the flavor and CS
brane with respect to background (\ref{eq:2}) and (\ref{eq:5}).
For flavor D7-brane, it is embedded at $x^{3},u=\mathrm{const}$ and
the onshell action and the holographic counterterm $S_{\mathrm{ct}}^{f}$
could be chosen as \cite{key-44,key-45,key-46},

\begin{align}
S_{f} & =S_{\mathrm{DBI}}+S_{\mathrm{ct}}^{f},\nonumber \\
S_{\mathrm{DBI}} & =-N_{f}T_{\mathrm{D}7}\int d^{8}xe^{\phi}\sqrt{g},\nonumber \\
S_{\mathrm{ct}}^{f} & =\frac{R}{3}T_{\mathrm{D7}}\int d^{7}x\sqrt{h_{\mathrm{D}7}}.\label{eq:58}
\end{align}
We note that the metric presented in (\ref{eq:56}) - (\ref{eq:58})
refers to the metric (\ref{eq:2}) and (\ref{eq:5}) in Einstein frame
which is defined as $g_{\mathrm{string}}=g_{\mathrm{Einstein}}e^{\phi/2}$.
For the CS brane, it is embedded at $x^{3}=\mathrm{const}$ and $r=r_{Q},r_{H}$
with respect to (\ref{eq:2}) and (\ref{eq:5}). The onshell action
of a CS brane evaluated in the confining background has been given
in (\ref{eq:40}) while it vanishes in the black brane background
(\ref{eq:2}) calculated by using (\ref{eq:58}). Therefore the total
onshell action including the bulk part, flavor part and CS part is,

\begin{equation}
S_{\mathrm{onshell}}=S_{\mathrm{bulk}}+S_{f}+S_{\mathrm{CS}}.
\end{equation}
Afterwards recalling the AdS/CFT dictionary,

\begin{equation}
\left\langle e^{-F}\right\rangle =e^{S_{\mathrm{onshell}}},
\end{equation}
with the solution (\ref{eq:2}) (\ref{eq:5}) we can obtain the free
energy $F$ respectively,

\begin{align}
F_{d} & =-\frac{1}{8}N_{c}^{2}\pi^{2}T^{4}V_{3}\beta_{T}-\frac{36\pi}{\lambda_{t}}N_{c}N_{\mathrm{D}}+\frac{1}{18}N_{c}N_{f}T^{3}V_{3}\lambda_{t}\left[\left(\pi+\ln64\right)q_{T}-2\right],\nonumber \\
F_{c} & =-\frac{M_{KK}^{4}N_{c}^{2}V_{3}\beta_{3}}{128\pi^{2}}-\frac{36\pi}{\lambda_{t}}N_{c}N_{\mathrm{D}}+\frac{\lambda M_{KK}^{3}N_{c}k_{0}V_{3}}{64\pi^{2}}G\left(q\right)+\frac{\lambda_{t}M_{KK}^{3}N_{c}N_{f}V_{3}\left[2+q\left(8+\pi-\ln4\right)\right]}{512\sqrt{2}\pi^{3/2}\Gamma\left(\frac{7}{4}\right)^{2}},\nonumber \\
\beta_{T} & =1/T,q_{T}=Q/r_{H}^{4},\beta_{3}=2\pi/M_{KK},\label{eq:61}
\end{align}
where $\lambda_{t}$ is the 't Hooft coupling constant defined as
$\lambda_{t}=g_{s}N_{c}$ and $F_{d,c}$ refers to the free energy
evaluated in the background (\ref{eq:2}) and (\ref{eq:5}) respectively.
$k_{0}$ is the number of CS brane and we have assumed $k_{0}$ and
$N_{f}$ take same order at large-$N_{c}$ due to the flavor and CS
branes as probes. $G\left(q\right)$ is a function defined as $G\left(q\right)=E_{\mathrm{CS}}\left(q\right)/E_{\mathrm{CS}}\left(0\right)$
whose behavior has been numerically illustrated in Figure \ref{fig:7}.
Follow the most discussion in gauge/gravity duality \cite{key-33,key-34,key-35,key-36},
the black brane (\ref{eq:2}) and soliton (bubble) solution (\ref{eq:5})
respectively corresponds to the deconfinement and confinement phase
in the dual theory, so the phase transition can be obtained by compare
their free energy which identifies the confinement/deconfinement phase
transition in the field theory as the Hawking-Page transition in the
bulk. According to (\ref{eq:61}), we can find the D-instantons as
D(-1)-branes negatively increase the bulk free energy as a contribution
of $\mathcal{O}\left(N_{c}^{2}\right)$ because at large-$N_{c}$
limit, $N_{d}/N_{c}$ must be fixed otherwise the backreaction of
the D-instantons in bulk vanishes. The critical temperature $T_{c}$
of the phase transition can be obtained by comparing the free energy
at $F_{d}=F_{c}$ which is evaluated as,

\begin{equation}
T_{c}=\frac{M_{KK}}{2\pi}-\frac{\lambda_{t}M_{KK}}{6\pi^{2}}\frac{k_{0}}{N_{c}}G\left(q\right)-\frac{\lambda_{t}M_{KK}}{432\pi^{2}}\frac{N_{f}}{N_{c}}\left(C_{1}+C_{2}q\right)+\mathcal{O}\left(N_{c}^{-2}\right),
\end{equation}
where $C_{1,2}$ are two constants given as,

\begin{align}
C_{1} & =64\pi+\frac{9\sqrt{2}\pi^{5/2}}{\Gamma\left(\frac{7}{4}\right)^{2}}\simeq464.66,\nonumber \\
C_{2} & =\frac{9\sqrt{2}\pi^{5/2}\left(8+\pi-\ln4\right)}{2\Gamma\left(\frac{7}{4}\right)^{2}}-32\pi\left(\pi+\ln64\right)\simeq551.815,\nonumber \\
q & =\frac{Q}{r_{KK}^{4}}=\frac{128\pi^{2}}{\lambda_{t}M_{KK}^{3}V_{3}}\frac{N_{\mathrm{D}}}{N_{c}}.
\end{align}
Notice in the large-$N_{c}$ limit, $q$ is fixed thus $G\left(q\right)$
is also fixed. So the critical temperature is not affected at $\mathcal{O}\left(N_{c}^{0}\right)$
while it decreases at $\mathcal{O}\left(N_{c}^{-1}\right)$ by the
presence of the D-instantons through the flavor and CS branes due
to $G\left(q\right)>0$. Since the behavior of $T_{c}$ is qualitatively
consistent with the behavior of $l_{c}\sim T^{-1}$ obtained by evaluating
the entanglement entropy, we may conclude that the entanglement entropy
is indeed able to characterize the deconfinement phase transition.

\section{Summary and discussion}

In this work, by compactifying on the supersymmetry breaking  $S^{1}$,
we construct the supergravity solution for $N_{c}$ black D3-branes
with dynamical $N_{\mathrm{D}}$ D-instantons, i.e. D(-1)-branes,
to obtain a 3d confining Yang-Mills in holography. To exhibit flavors
and the CS term in the dual theory, we also add flavor and CS branes
as probe into the bulk geometry hence the dual theory is expected
to be a 3d YMCS with matters or CS QCD-like theory. The low-energy
regime of the 3d dual theory is analyzed by the IIB supergravity solution
which geometrically shows the spontaneous breaking of the chiral symmetry
$U\left(N_{f}\right)$ down to $U\left(p\right)\times U\left(N_{f}-p\right)$,
$p\in\mathbb{Z}$. And at very low-energy, D-instantons could reduce
to a pure CS theory. Due to the presence of the dynamical D-instantons,
the embedding function of the flavor branes depends on the non-zero
charge density of the D-instantons which is realized to be metastable
vacua of instantons in the dual theory. Then we further evaluate the
vacuum structure of the dual theory by including both flavor and CS
branes which leads to a topological phase transition determined by
the order parameter $m^{*}$ in the large-$N_{c}$ limit and $m^{*}$
is increased by the presence of D-instantons as it is expected. This
behavior of $m^{*}$ can be interpreted as the topological contribution
from the CS term in the dual theory, similarly as the topological
contribution to mass in the CS theory. Moreover, we additionally evaluate
the entanglement entropy and total free energy in holography to investigate
the critical length $l_{c}$ and critical temperature $T_{c}$ which
is expected to be the characters of the deconfinement phase transition.
The behavior of $T_{c}$ is in qualitative agreement with the behavior
of $l_{c}$ which implies the quantum entanglement entropy could indeed
be a tool to determine confinement/deconfinement in this holographic
approach.

We would like to give some comments to close this work. First, we
notice that the discrepancy between topological phases characterized
by $m^{*}$ becomes vanished if $E_{\mathrm{CS}}\rightarrow0$. And
in the black brane background (\ref{eq:2}), the CS brane is excepted
to be embedded at $r=r_{H}$ to minimize its energy which leads to
a vanished $E_{\mathrm{CS}}$. Since the black brane background corresponds
to a dual theory at finite temperature, the topological structure
of the vacuum may therefore becomes vanished. So in this sense, our
model might provide a holographic interpretation of that why the topological
aspects of hot QCD by instantons is quite difficult to be measured
in experiment \cite{key-47,key-48,key-49,key-50}. 

Second, it is expected the topological phase transition is second
order \cite{key-1,key-2,key-3,key-4,key-5,key-6,key-7} if the number
of the CS brane is $\mathcal{O}\left(N_{c}\right)$. This can be achieved
by taking into account the backreaction of CS branes. However the
number of CS brane is given by $\int_{S^{1}}F_{1}$ which relates
to the boundary value of $C_{0}$ in our current setup. So the bulk
dynamic could not involve the backreaction of CS branes in this work.
The valid way to include the backreaction of CS brane is to solve
the IIB supergravity action with a fluctuation of $C_{0}$ sourced
by the CS branes then the next-to-the-leading-order contribution in
the large-$N_{c}$ limit to the vacuum structure would be able to
analyze in this sense. However, we would like to leave this for the
future study.

Last but not least, since the topological entanglement entropy is
defined as the finite part of the entanglement \cite{key-51,key-52}
which could be the measure of the topological order, $\Delta S$ should
relate to the topological entanglement entropy. So the critical length
$l_{c}$ seemingly shows the transition between the phases with different
topological entanglement entropy. Thus if the entanglement entropy
can characterize the deconfinement phase transition, $T_{c}$ may
also reflect some properties of the topological order in the theory.
However our result also shows, in the large-$N_{c}$ limit, $T_{c}$
becomes nearly independent on the instantons while the behavior of
$l_{c}$ remains to be determined by the instantons. Accordingly it
seems the entanglement entropy is more sensitive to the topological
properties of the theory than the critical temperature. And we expect
it could be an instructive way to study the topological structure
of YMCS theory.

\section*{Acknowledgements}

This work is supported by the National Natural Science Foundation
of China (NSFC) under Grant No. 12005033, the research startup foundation
of Dalian Maritime University in 2019 under Grant No. 02502608 and
the Fundamental Research Funds for the Central Universities under
Grant No. 3132021205.


\begin{thebibliography}{10}
\bibitem{key-1} S. Giombi, S. Minwalla, S. Prakash, S. P. Trivedi,
S. R. Wadia and X. Yin, \textquotedblleft Chern- Simons Theory with
Vector Fermion Matter\textquotedblright , Eur. Phys. J. C 72, 2112
(2012), arXiv: 1110.4386.

\bibitem{key-2} O. Aharony, G. Gur-Ari and R. Yacoby, \textquotedblleft d=3
Bosonic Vector Models Coupled to Chern- Simons Gauge Theories\textquotedblright ,
JHEP 1203 (2012) 037, arXiv: 1110.4382.

\bibitem{key-3} C. -M. Chang, S. Minwalla, T. Sharma and X. Yin,
\textquotedblleft ABJ Triality: from Higher Spin Fields to Strings\textquotedblright ,
J.Phys.A 46 (2013) 214009, arXiv: 1207.4485.

\bibitem{key-4} S. Jain, S. P. Trivedi, S. R. Wadia and S. Yokoyama,
\textquotedblleft Supersymmetric Chern-Simons Theories with Vector
Matter\textquotedblright , JHEP 10 (2012) 194, arXiv: 1207.4750.

\bibitem{key-5} O. Aharony, G. Gur-Ari and R. Yacoby, \textquotedblleft Correlation
Functions of Large N Chern- Simons-Matter Theories and Bosonization
in Three Dimensions\textquotedblright , JHEP 02 (2013), arXiv: 1207.4593.

\bibitem{key-6} G. Gur-Ari and R. Yacoby, \textquotedblleft Correlators
of Large N Fermionic Chern-Simons Vector Models\textquotedblright ,
JHEP 02 (2013) 150, arXiv: 1211.1866.

\bibitem{key-7} S. Jain, S. Minwalla and S. Yokoyama, ``Chern Simons
duality with a fundamental boson and fermion'', JHEP 1311 (2013)
037, arXiv: 1305.7235.

\bibitem{key-8} O. Aharony, ``Baryons, monopoles and dualities in
Chern-Simons-matter theories'', JHEP 1602 (2016) 093, arXiv: 1512.00161. 

\bibitem{key-9} P.S. Hsin and N. Seiberg, ``Level/rank Duality and
Chern-Simons-Matter Theories'', JHEP 1609 (2016) 095, arXiv: 1607.07457.

\bibitem{key-10} O. Aharony, S. S. Gubser, J. M. Maldacena, H. Ooguri
and Y. Oz, ``Large N field theories, string theory and gravity'',
Phys. Rept. 323 (2000) 183, arXiv: hep-th/9905111.

\bibitem{key-11} E. Witten, ``Anti-de Sitter space and holography'',
Adv.Theor.Math.Phys. 2 (1998) 253-291, arXiv: hep-th/9802150.

\bibitem{key-12} K. Becker, M. Becker, J.H. Schwarz, ``String Theory
and M-Theory, A Modern Introduction'', Cambridge University Press,
Cambridge, 2007.

\bibitem{key-13} E. Witten, ``Anti-de Sitter space, thermal phase
transition, and confinement in gauge theories'', Adv. Theor. Math.
Phys. 2 (1998), 505-532, arXiv: hep-th/9803131.

\bibitem{key-14} D. K. Hong and H. U. Yee, ``Holographic aspects
of three dimensional QCD from string theory'', JHEP 1005 (2010) 036,
Erratum: JHEP 1008 (2010) 120, arXiv:1003.1306.

\bibitem{key-15} R. Argurio, A. Armoni, M. Bertolini, F. Mignosa,
P. Niro, ``Vacuum structure of large $N$ QCD\textsubscript{3} from
holography'', JHEP 07 (2020) 134, arXiv: 2006.01755.

\bibitem{key-16} M. Fujita, W. Li, S. Ryu and T. Takayanagi, ``Fractional
Quantum Hall Effect via Holography: Chern-Simons, Edge States, and
Hierarchy'', JHEP 06 (2009), 066, arXiv: 0901.0924.

\bibitem{key-17} T. Schäfer, E. V. Shuryak, ``Instantons in QCD'',
Rev.Mod.Phys. 70 (1998) 323-426, arXiv: hep-ph/961045.

\bibitem{key-18} D. J. Gross, R. D. Pisarski, L. G. Yaffe, ``QCD
and Instantons at Finite Temperature'', Rev.Mod.Phys. 53 (1981) 43.

\bibitem{a1} E. Witten, ``Small instantons in string theory'',
Nucl.Phys.B 460 (1996) 541-559, arXiv: hep-th/9511030.

\bibitem{a2} M. R. Douglas, ``Branes within branes'', NATO Sci.Ser.C
520 (1999) 267-275, arXiv: hep-th/9512077.

\bibitem{key-19} G. W. Gibbons, M. B. Green, M. J. Perry, \textquotedblleft Instantons
and seven-branes in type IIB superstring theory\textquotedblright ,
Phys.Lett.B 370 (1996) 37-44, arXiv: hep-th/9511080.

\bibitem{key-20} H. Liu, A. A. Tseytlin, ``D3-brane D instanton
configuration and N=4 superYM theory in constant selfdual background'',
Nucl.Phys.B 553 (1999) 231-249, arXiv: hep-th/9903091.

\bibitem{key-21} A. Kehagias, ``On asymptotic freedom and confinement
from type IIB supergravity'', Phys.Lett.B 456 (1999) 22-27, arXiv:
hep-th/9903109.

\bibitem{key-22} B. Gwak, M. Kim, B. H. Lee, Y. Seo, S.J. Sin, ``Holographic
D Instanton Liquid and chiral transition'', Phys.Rev.D 86 (2012)
026010, arXiv: 1203.4883.

\bibitem{key-23} S. Li, S. Lin, ``D-instantons in Real Time Dynamics'',
Phys.Rev.D 98 (2018) 6, 066002, arXiv: 1711.06365.

\bibitem{key-24} Z. Q. Zhang, D. F. Hou, G. Chen, ``Imaginary potential
of moving quarkonia in a D-instanton background'', J.Phys.G 44 (2017)
11, 115001, arXiv: 1710.06579.

\bibitem{key-25} S. Li, ``Holographic Schwinger effect in the confining
background with D-instanton'', arXiv: 2005.11955.

\bibitem{key-26} A. Karch and E. Katz, ``Adding flavor to AdS /
CFT'', JHEP 0206 (2002) 043, arXiv: hep- th/0205236.

\bibitem{key-27} A. Baumgartner, ``Flavor broken QCD\textsubscript{3}
at large N'', JHEP 08 (2020) 145, arXiv: 2005.11339.

\bibitem{key-28} G. V. Dunne, ``Aspects of Chern-Simons theory'',
arXiv: hep-th/9902115.

\bibitem{key-29} S. Li, T. Jia, ``Matrix model and Holographic Baryons
in the D0-D4 background'' , Phys.Rev.D 92 (2015) 4, 046007, arXiv:
1506.00068.

\bibitem{key-30} C. Wu, Z.G Xiao, D. Zhou, ``Sakai-Sugimoto model
in D0-D4 background'', Phys.Rev.D 88 (2013) 2, 026016, arXiv: 1304.2111.

\bibitem{key-31} F. Bigazzi, A. L. Cotrone, R. Sisca, ``Notes on
Theta Dependence in Holographic Yang-Mills'', JHEP 08 (2015) 090,
arXiv: 1506.03826. 

\bibitem{key-32} S. Li, ``A holographic description of theta-dependent
Yang-Mills theory at finite temperature'', Chin.Phys.C 44 (2020)
1, 013103, arXiv: 1907.10277.

\bibitem{key-33} O. Aharony, J. Sonnenschein, S. Yankielowicz, ``A
Holographic model of deconfinement and chiral symmetry restoration'',
Annals Phys. 322 (2007) 1420-1443, arXiv: hep-th/0604161.

\bibitem{key-34} S. Li, T. Jia, ``Dynamically flavored description
of holographic QCD in the presence of a magnetic field'', Phys.Rev.D
96 (2017) 6, 066032, arXiv: 1604.07197.

\bibitem{key-35} S. Li, A. Schmitt, Qun Wang, ``From holography
towards real-world nuclear matter'', Phys.Rev.D 92 (2015) 2, 026006,
arXiv: 1505.04886.

\bibitem{key-36} F. Bigazzi, A. L. Cotrone, ``Holographic QCD with
Dynamical Flavors'', JHEP 01 (2015) 104, arXiv: 1410.2443.

\bibitem{key-37} I. R. Klebanov, D. Kutasov, A. Murugan, ``Entanglement
as a probe of confinement'', Nucl.Phys.B 796 (2008) 274-293, arXiv:
0709.2140.

\bibitem{key-38} J. Knaute, B. Kämpfer, ``Holographic Entanglement
Entropy in the QCD Phase Diagram with a Critical Point'', Phys.Rev.D
96 (2017) 10, 106003, arXiv: 1706.02647.

\bibitem{key-39} N. Jokela, J. G. Subils, ``Is entanglement a probe
of confinement?'', JHEP 02 (2021) 147, arXiv: 2010.09392.

\bibitem{key-40} M. A. Akbari, M Lezgi, ``Holographic QCD, entanglement
entropy, and critical temperature'', Phys.Rev.D 96 (2017) 8, 086014,
arXiv: 1706.04335.

\bibitem{key-41} S. Ryu, T. Takayanagi, ``Holographic derivation
of entanglement entropy from AdS/CFT'', Phys.Rev.Lett. 96 (2006)
181602, arXiv: hep-th/0603001.

\bibitem{key-42} A. W. Peet and J. Polchinski, ``UV / IR relations
in AdS dynamics'', Phys. Rev. D 59 (1999), 065011, arXiv:hep-th/9809022. 

\bibitem{key-43} R. Emparan, C.V. Johnson and R.C. Myers, \textquotedblleft Surface
terms as counterterms in the AdS/CFT correspondence\textquotedblright ,
Phys. Rev. D 60 (1999) 104001, arXiv:hep-th/9903238.

\bibitem{key-44} D. Mateos, R. C. Myers and R. M. Thomson, \textquotedblleft Thermodynamics
of the brane\textquotedblright , JHEP 0705, 067 (2007), arXiv: hep-th/0701132.

\bibitem{key-45} P. Benincasa, \textquotedblleft A note on Holographic
Renormalization of Probe D-Branes\textquotedblright , arXiv:0903.4356.

\bibitem{key-46} I. Papadimitriou, \textquotedblleft Holographic
Renormalization of general dilaton-axion gravity\textquotedblright ,
JHEP 1108, 119 (2011), arXiv:1106.4826.

\bibitem{key-47} D. Kharzeev, R.D. Pisarski, M. H. G. Tytgat, ``Possibility
of spontaneous parity violation in hot QCD'', Phys.Rev.Lett. 81 (1998)
512-515, arXiv: hep-ph/9804221.

\bibitem{key-48} K. Buckley, T. Fugleberg, A. Zhitnitsky, ``Can
theta vacua be created in heavy ion collisions?'', Phys.Rev.Lett.
84 (2000) 4814-4817, arXiv: hep-ph/9910229.

\bibitem{key-49} D. E. Kharzeev, L. D. McLerran, H. J. Warringa,
``The Effects of topological charge change in heavy ion collisions:
'Event by event P and CP violation'{}'', Nucl.Phys.A 803 (2008) 227-253,
arXiv: 0711.0950.

\bibitem{key-50} D. Kharzeev, ``Parity violation in hot QCD: Why
it can happen, and how to look for it'', Phys.Lett.B 633 (2006) 260-264,
arXiv: hep-ph/0406125.

\bibitem{key-51} A. Kitaev and J. Preskill, \textquotedblleft Topological
entanglement entropy\textquotedblright , Phys. Rev. Lett. 96 (2006)
110404, arXiv:hep-th/0510092.

\bibitem{key-52} M. Levin, X. G. Wen, \textquotedblleft Detecting
topological order in a ground state wave function\textquotedblright ,
Phys. Rev. Lett. 96 (2006) 110405, arXiv:cond-mat/0510613.
\end{thebibliography}
\end{document}